\documentclass{aastex}
\usepackage{spr-astr-addons}
\usepackage{url}
\usepackage[utf8]{inputenc}

\RequirePackage{color}

\begin{document}

\title{Compact stellar model in presence of pressure anisotropy in modified Finch Skea spacetime}
\shorttitle{Compact stellar model in presence of pressure anisotropy in modified Finch Skea spacetime}

\shortauthors{Piyali Bhar \& Pramit Rej }

\author{Piyali Bhar }

\altaffiltext{}{ Department of
Mathematics, Government General Degree College, Singur, West Bengal 712 409,
India}
\altaffiltext{}{Email:piyalibhar90@gmail.com}

\author{Pramit Rej}

\altaffiltext{}{Department of
Mathematics, Sarat Centenary College, Dhaniakhali, Hooghly, West Bengal 712 302, India }
\altaffiltext{}{Email: pramitrej@gmail.com}

\begin{abstract}
A new model of anisotropic compact star is obtained in our present paper by assuming the pressure anisotropy. The proposed model is singularity free. The model is obtained by considering a physically reasonable choice for the metric potential $g_{rr}$ which depends on a dimensionless parameter `n'. The effect of $n$ is discussed numerically, analytically and through plotting. We have concentrated a wide range for n ($10\leq n \leq 1000$) for drawing the profiles of different physical parameters. The maximum allowable mass for different values of $n$ have been obtained by M-R plot. We have checked that the stability of the model is increased for larger value of $n$. For the viability of the model we have considered two compact stars PSR J1614-2230 and EXO 1785-248. We have shown that the expressions for the anisotropy factor and the metric component may serve as generating functions for uncharged stellar models in the context of the general theory of relativity.
\end{abstract}


\maketitle

\section{Introduction}
The term `compact object' is mainly used in astronomy to describe collectively to white dwarfs, neutron stars and black hole. It is well known that stars are an isolated body that is bounded by self-gravity, and which radiates energy supplied by an internal source. Most compact objects are formed to a point of the stellar evolution when the internal radiation pressure from the nuclear fusions of a star cannot balance the external gravitational force and the star collapses under its own weight. It is familiar that the model of compact star can be obtained by solving Einstein's field equations in the context of General theory of relativity. There are large numbers of papers on exact solution of Einstein's field equations for spherically symmetric perfect fluid spheres \cite{39,40}. \cite{41} have obtained two new classes of solutions of field equations with constant proper mass densities. \cite{42} has solved field equations for finding interior solutions for spherically symmetric, static and conformal flat anisotropic fluid spheres. According to \cite{43} the pressure inside the highly compact astrophysical objects like an X-ray pulsar, Her-X-1, X-ray buster 4U 1820-30, the millisecond pulsar PSR J1614-2230, LMC X-4 etc. that have a core density beyond the nuclear density ($10^{15}$ gm/cc) show anisotropic nature, i.e., the pressure inside these compact objects can be decomposed into two parts: the radial pressure $p_r$ and the transverse pressure $p_t$. Existence of a solid stellar core, the presence of a type-3A superfluid,
pion condensation, different kinds of phase
transitions, mixture of two gases etc. are reasonable for pressure anisotropy \cite{72,73,74,71}. A large number of works have been done by assuming pressure anisotropy \cite{471,481,451,441,461,491}.

To obtain the maximum value of the ratio of mass to radius for a model of compact star is an important problem in relativistic astrophysics since
``the existence of such a bound is intriguing because it occurs well before the appearance
of an apparent horizon at $M = R/2$" as proposed by \cite{guven}. The authors have investigated the upper limit of $M/R$ for compact
general relativistic configurations by assuming that inside the star
the radial stress $p_r$ is different from the tangential pressure $p_t$. It was also investigated by them that if the density is monotonically
decreasing and radial pressure is greater than the tangential pressure then the upper bound $8/9$ is still valid to the entire bulk if $m$ is
replaced by the quasi-local mass. Several bounds on the mass-radius ratio and anisotropy parameter have also been found, for models in which the anisotropic factor is proportional to $r^2$ \cite{19}. Bounds on $m/r$ for static objects with a positive cosmological constant was obtained by \cite{andboh09}. According to \cite{hh}, for arbitrarily large anisotropy, in principle there is neither limiting mass nor limiting redshift and they also proposed that Semi-realistic equations of state lead to a mass of $3-4 M_{\odot}$ for neutron stars with
an anisotropic equation of state. \cite{liang1} have analytically obtained the maximum equilibrium mass and surface redshift in the case of incompressible neutron matter. For a relativistic stellar model the value of the redshift can not be arbitrarily large. \cite{bondi92} considered the relation between redshift and the ratio of the trace of the pressure tensor to local density and proved that when anisotropic pressures are allowed, considerably larger redshift values can be obtained. \cite{iva1} obtained the maximum value of the redshift for anisotropic stars. For realistic anisotropic star models, the surface redshift cannot exceed the values $3.842$ or $5.211$ when the tangential pressure respectively satisfies the strong or the dominant energy condition, whereas this bound in the perfect fluid case is $2$.

The dynamical stability of spherically symmetric gravitational equilibria of cold stellar objects made of bosons and fermions was proposed by \cite{jet2}. Using the Einstein-Klein-Gordon equation, Jetzer et al. explored the dynamical instability of the static real scalar field \cite{jet}. \cite{pmn15} studied the behavior of static spherically symmetric relativistic objects with locally anisotropic matter distribution considering the Tolman VII form for the gravitational potential $g_{rr}$ in curvature coordinates together with the linear relation between the energy density and the radial pressure. \cite{bhar15d} proposed a model of superdense star admitting conformal motion in presence of quintessence field by taking \cite{vt82} ansatz motivated by accelerating phase of our present universe. Ivanov proposed a physically realistic stellar model with a simple expression for the energy density with conformally flat
interior and discussed all the physical properties without graphic proofs \cite{ivanov19}. Recently, \cite{ranjan} obtained a new class of solutions by revisiting the Vaidya-Tikekar stellar model in the linear regime and discussed the impact of the curvature parameter $K$ of the Vaidya-Tikekar model, which characterizes a departure from homogeneous spherical distribution, on the mass-radius relationship of the star. \cite{musta} considered spherically symmetric space-time with an anisotropic fluid distribution. In particular, they used the Karmarkar condition to explore the compact star solutions.  \\
To study the model of compact star by utilizing \cite{fs89} metric potential is a very interesting platform to the researchers. \cite{hm14} obtained charged Finch-Skea stars and these models are given in terms of Bessel functions and obey a barotropic equation of state. \cite{sr} obtained the model of Stars with a quadratic equation
of state with the Finch-Skea geometry and this work was extended by Pandya et al. for a generalized form of
the gravitational potential \cite{panda}. \cite{kalam}  proposed quintessence stars with both
dark energy and anisotropic pressures. Charged anisotropy stellar model in the background of Finch and Skea geometry has been obtained by \cite{ma12} and the exact solutions
have be expressed in terms of elementary functions, Bessel functions and modified Bessel
functions. \cite{pb15} proposed a new model
of an anisotropic strange star which admits the Chaplygin
equation of state. The study of compact star model in the context of Finch and Skea geometry has been
studied in matter distributions with lower dimensions also \cite{pb14,ayan}. Relativistic solutions of anisotropic charged compact objects
in hydrodynamical equilibrium with Finch-Skea geometry in the usual four and in
higher dimensions has been studied by \cite{dey}.\par

Inspired by all of these earlier works, in the present paper, we obtain a new model of singularity free compact star by assuming a physically reasonable anisotropic factor along with the metric potential $g_{rr}.$
Our paper is arranged as follows: In sect.~\ref{sec2}~ the basic field equations have been discussed. In next sect., we have described about the metric potential $g_{rr}$. sect.~\ref{sec4} is devoted to the description of the new model. In sect.~\ref{bou}, we matched our interior solution to the exterior Schwarzschild line element. Physical analysis of the obtained model is described in sect.~\ref{pa}. In next section, the stability condition of the present model has been discussed and finally some comparative study of the present model for different values of $n$ is given in sect.~\ref{dis}
\section{Interior Spacetime and Einstein field Equations}\label{sec2}
The structure of compact and massive stars is determined by Einstein's field equations,
\begin{eqnarray}
G_{\mu\nu}&=&\kappa\frac{G}{c^4} T_{\mu \nu},\label{e11}
\end{eqnarray}
$G_{\mu\nu},\,T_{\mu\nu}$ are Einstein tensor and energy-momentum tensor respectively. $G$ and $c$ respectively denote the universal gravitational constant and speed of the light.\par
A non-rotating spherically symmetry $4$-D spacetime in Schwarzschild coordinates $x^{\mu}=(t,\,r,\,\theta,\,\phi)$ is described by the line element,
\begin{equation}
ds^{2}=-A^{2}dt^{2}+B^{2} dr^{2}+r^{2}(d\theta^{2}+\sin^{2}\theta d\phi^{2}),
\end{equation}
Where $A$ and $B$ are static, i.e., functions of the radial coordinate `$r$' only and these are called the gravitational potentials. They satisfy the Einstein's field equation (\ref{e11}).\\
We also assume that, inside the stellar interior, the matter is anisotropic in nature and therefore, we write the corresponding energy-momentum tensor as,
\begin{equation}
T_{\nu}^{\mu}=(\rho+p_r)u^{\mu}u_{\nu}-p_t g_{\nu}^{\mu}+(p_r-p_t)v^{\mu}v_{\nu}
\end{equation}
with $ u^{i}u_{j} =-v^{i}v_j = 1 $ and $u^{i}v_j= 0$. Here the vector $v^{i}$ is the space-like vector and $u_i$ is the fluid 4-velocity and it is orthogonal to $v^{i}$, $\rho$ is the matter density, $p_t$ and $p_r$ are respectively the transverse and radial pressure of the fluid and these two pressure components act in the perpendicular direction to each other.\\
The Einstein field equations assuming $G=1=c$ are given by
\begin{eqnarray}
\kappa \rho&=&\frac{1}{r^2}\left(1-\frac{1}{B^2}\right)+\frac{2B'}{B^3r},\label{1a}\\
\kappa p_r &=&\frac{1}{B^2}\left(\frac{1}{r^2}+\frac{2A'}{Ar}\right)-\frac{1}{r^2},\label{2a}\\
\kappa p_t &=&\frac{A''}{AB^2}-\frac{A'B'}{AB^3}+\frac{1}{B^3rA}(A'B-B'A).\label{3a}
\end{eqnarray}
The mass function for our present model is obtained as,
\begin{eqnarray}
m(r)&=&4\pi\int_0^{r}\omega^{2}\rho(\omega)d\omega,\label{4}
\end{eqnarray}
Using (\ref{1a})-(\ref{3a}) we get,
\begin{eqnarray}
\frac{2A'}{A}&=&\frac{\kappa r p_r+\frac{2m}{r^2}}{1-\frac{2m}{r}},\label{k9}\\
\frac{dp_r}{dr}&=&-(\rho+p_r)\frac{A'}{A}+\frac{2}{r}(p_t-p_r),\label{k10}
\end{eqnarray}
Combining (\ref{k9}) and (\ref{k10}), one can finally obtain :
\begin{eqnarray}
\frac{dp_r}{dr}&=&-\frac{\rho+p_r}{2}\frac{\left(\kappa r p_r+\frac{2m}{r^2}\right)}{\left(1-\frac{2m}{r}\right)}+\frac{2}{r}(p_t-p_r).\label{k11}
\end{eqnarray}
The eqn.~(\ref{k11}) is called the Tolman-Oppenheimer-
Volkoff (TOV) equation of a hydrostatic equilibrium
for the anisotropic stellar configuration and $\kappa = 8\pi$ being the Einstein's constant.\par

Where `prime' indicates differentiation with respect to radial co-ordinate $r$.
Our goal is to solve the eqns. (\ref{1a})-(\ref{3a}).
\section{Choice of the metric potential}\label{sec3}
In the system (\ref{1a})-(\ref{3a}), we have three equations with five unknown ($\rho,\,p_r,\,p_t,\,A,\,B$). To solve this system we are free to choose any two of them. Now by our previous knowledge of algebra we know that there are ${5 \choose 2 }=10$ possible ways to chose any two unknowns. \cite{ratanpal}, \cite{pb1}, \cite{pb2} choose $B^2$ along with $p_r$, \cite{bhra} choose $\rho$ along with $p_r$, \cite{murad}, \cite{thi} choose $A^2$ with $\Delta$ to model different compact stars. But a very popular technique is to choose $B^2$ along with an equation of state (EoS), i.e., a relation between the matter density $\rho$ and radial pressure $p_r$. Several papers were published in this direction \cite{a1,a2,a3,a5,a7,a8,a9,a4}.\\
To solve the equation (\ref{1a})-(\ref{3a}) let us assume the expression of the co-efficient of $g_{rr}$, i.e., $B^2$ as,
\begin{eqnarray}\label{enu1}
 B^2&=&\left(1 + \frac{r^2}{R^2}\right)^n,
\end{eqnarray}
where $n\neq 0 $ is a real number. If one takes $n=0$, then from eqn. (\ref{1a}), we get $\rho=0$, which is not physically reasonable.
The gravitational potential $B^2$ is well behaved and finite at the origin. It is also continuous in the interior and it is
important to realize that this choice for $B^2$ is physically reasonable. For $n=1$, the metric potential reduces to well known \cite{fs89} potential. The metric potential in (\ref{enu1}) was used earlier by \cite{panda} to model a compact star by using a proper choice of the radial pressure $p_r$ and they proved that the model is compatible with
observational data.\\
Assuming $G=c=1$ and by using (\ref{enu1}) into (\ref{1a}), the expression for matter density is obtained as,
\begin{eqnarray}
\kappa \rho&=&\frac{1 - \left(1 + \frac{r^2}{R^2}\right)^{-n}}{r^2} + \frac{2 n \left(1 + \frac{r^2}{R^2}\right)^{-1 - n}}{R^2}.\label{rho1}
\end{eqnarray}

\section{Proposed model}\label{sec4}
Using the expression of $B^2$, eqns. (\ref{2a}) and (\ref{3a}) becomes,
\begin{eqnarray}
\kappa p_r &=&\left(1 + \frac{r^2}{R^2}\right)^{-n}\left(\frac{1}{r^2}+\frac{2A'}{Ar}\right)-\frac{1}{r^2},\label{4a}\\
\kappa p_t &=&\frac{A''}{A}\left(1 + \frac{r^2}{R^2}\right)^{-n}-\frac{A'}{A}\frac{n r \left(1 + \frac{r^2}{R^2}\right)^{-1 - n}}{R^2}\nonumber\\&&+\frac{1}{rA}\Big\{A'-\frac{Anr}{R^2}  \left(1 + \frac{r^2}{R^2}\right)^{-1}\Big\}\left(1 + \frac{r^2}{R^2}\right)^{-n}.\label{5a}
\end{eqnarray}
From equation (\ref{4a}) and (\ref{5a}), it is clear that once we have the expression for $A$, one can obtain the expressions for $~p_r,~p_t$ in closed form. Instead for a choice for $A$, for our present model we introduce the anisotropic factor $\Delta$, the difference between these two pressures, i.e., $p_t-p_r$. This anisotropic factor measures the anisotropy inside the stellar interior and it creates an anisotropic force which is defined as $\frac{2\Delta}{r}$. This force may be positive or negative by depending on the sign of $\Delta$, but at the center of the star, the force is zero since the anisotropic factor vanishes there.\\

\begin{figure}[htbp]
    \centering
        \includegraphics[scale=.6]{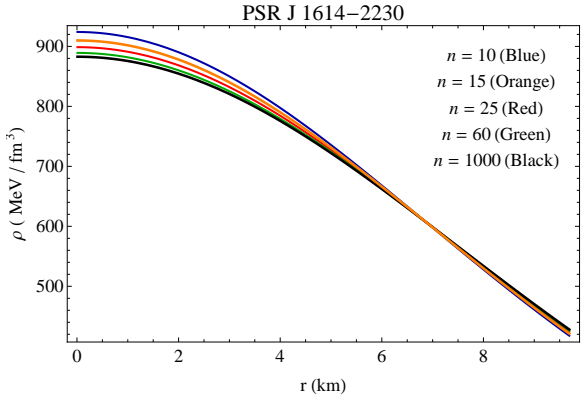}
       \caption{Matter density $\rho$ is plotted against $r$ inside the stellar interior for the compact star PSR J1614-2230 for different values of `n' mentioned in the figure.}\label{rho}
\end{figure}

From (\ref{4a}) and (\ref{5a})we get,
\begin{eqnarray}\label{10}
\frac{A''}{A}-\frac{A'}{A}\frac{1}{r}\left(1+\frac{n r^2}{r^2 + R^2}\right)+\frac{-1 + (1 + \frac{r^2}{R^2})^n}{r^2} \nonumber\\=\frac{n}{r^2 + R^2}+\kappa \left(1+\frac{r^2}{R^2}\right)^n\Delta,
\end{eqnarray}

To solve the equations (\ref{4a}) and (\ref{5a}), we choose, the expression of $\Delta$ as,
\begin{eqnarray}
\kappa \Delta &=& \frac{1}{r^2} - \frac{\left(1 + \frac{r^2}{R^2}\right)^{-n} \Big((1 + n) r^2 + R^2\Big)}{r^2(r^2 + R^2)},\label{6a}
\end{eqnarray}
 Whenever we choose the expression one has to remember that the expression of $\Delta$ is to be chosen in such a manner that (i) it should vanish at the center of the star, (ii) it does not suffer from any kind of singularities , (iii) $\Delta$ is positive inside the stellar interior and finally (iv) the field equation can be integrated easily with this choice of $\Delta$. To obtain the model of compact star, \cite{dey}, \cite{maharaj1}, \cite{murad} choose a physically reasonable choice of $\Delta$  to find the exact solutions for the Einstein-Maxwell equations. Our present choice of the anisotropic factor satisfies the above conditions which will be verified in the coming section and therefore this choice is physically reasonable.\par
 \begin{figure}[htbp]
    \centering
        \includegraphics[scale=.6]{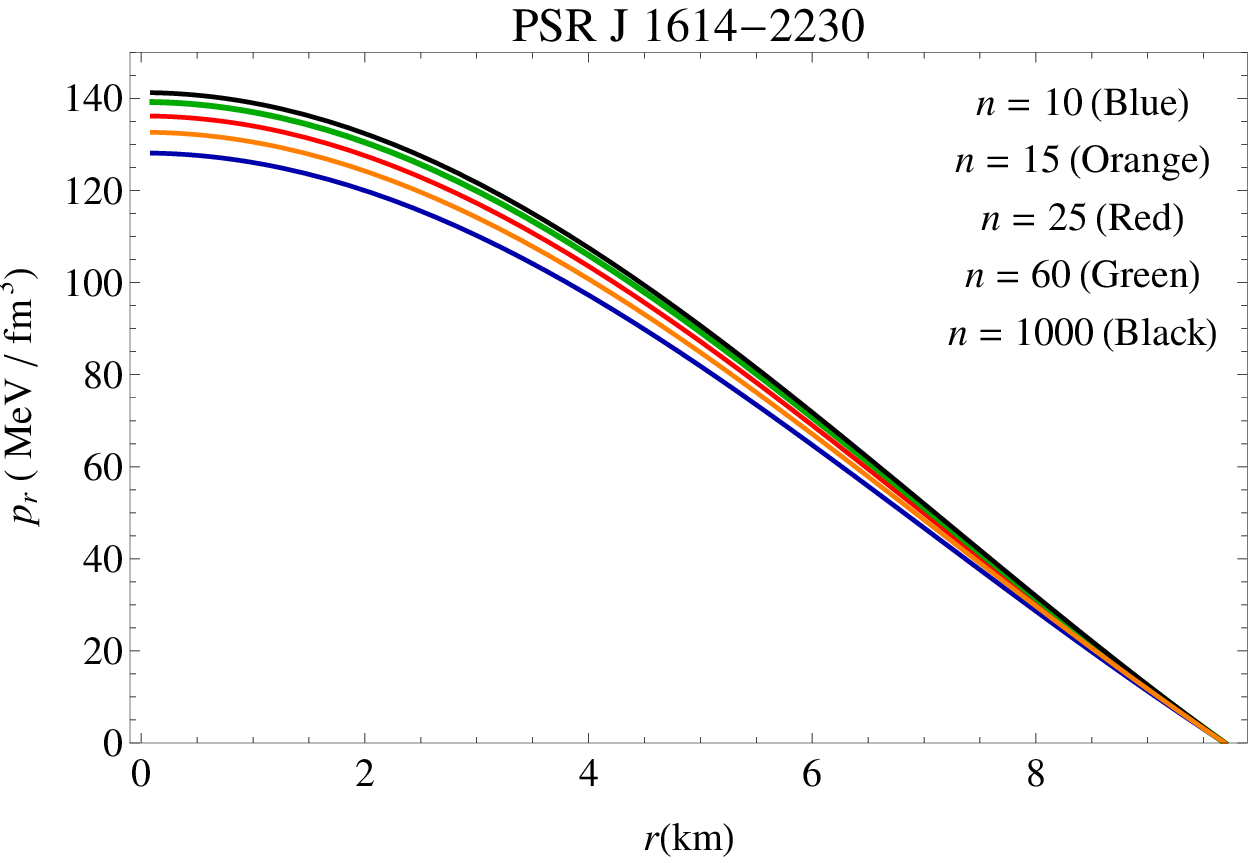}
         \includegraphics[scale=.6]{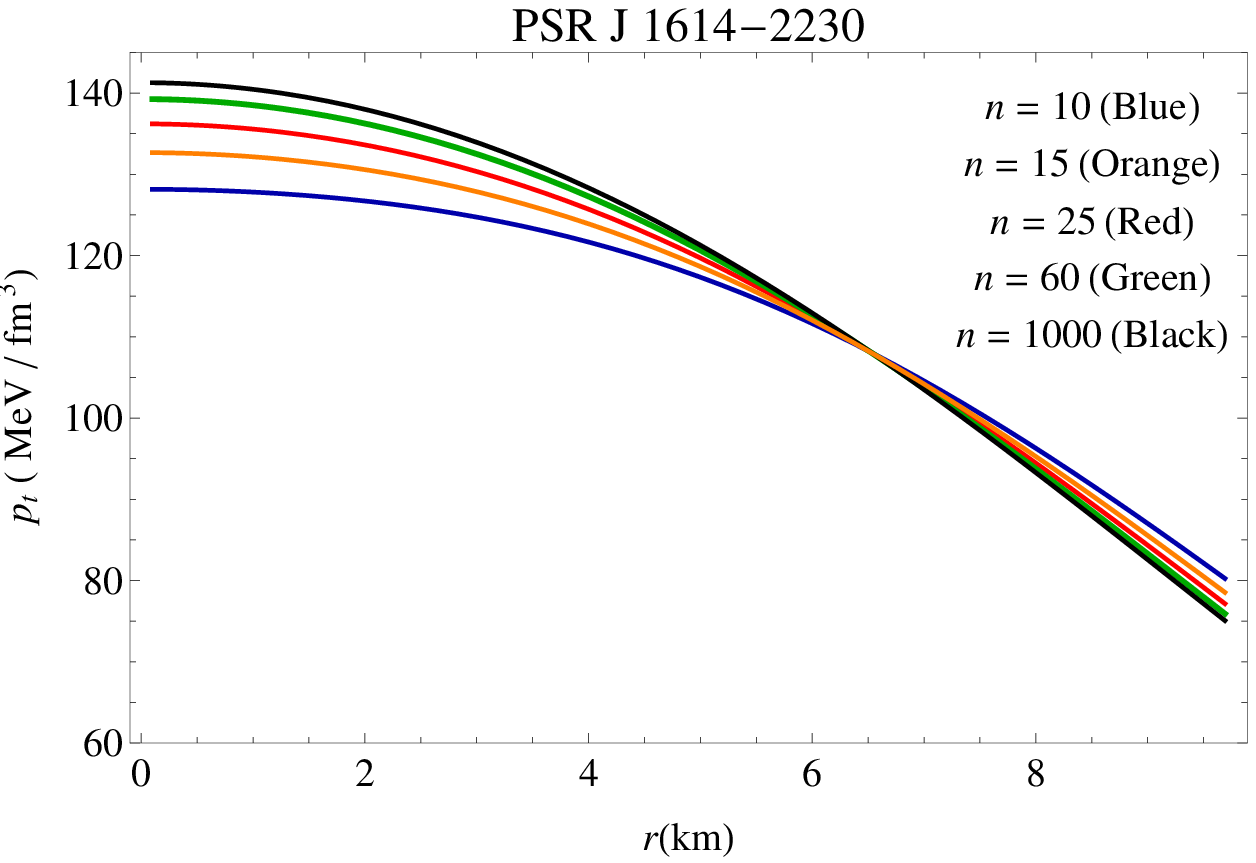}
       \caption{(Top) Radial pressure $p_r$ and (bottom) transverse pressure $p_t$ are plotted against $r$ inside the stellar interior for the compact star PSR J1614-2230 for different values of `n' mentioned in the figures.\label{pr}}
\end{figure}
Solving (\ref{4a}) and (\ref{5a}) with the help of (\ref{6a}), we get the following equation:
\begin{eqnarray}
A''=A' \left(1 + \frac{n r^2}{r^2 + R^2}\right) \frac{1}{r},\label{1n}
\end{eqnarray}
Solving (\ref{1n}), we obtain the another metric coefficient as,
\begin{eqnarray}\label{elambda1}
A^2&=&\left(D + \frac{C (r^2 + R^2)^{1 + \frac{n}{2}}}{2 + n}\right)^2,
\end{eqnarray}
Where $C,\,D$ are constants of integrations which can be obtained from the matching conditions.
Using the expression for $A^2$, the expressions for radial and transverse pressure are obtained as,
\begin{eqnarray}
\kappa p_r &=& \frac{
 2 C (2 + n) \left(1 + \frac{r^2}{R^2}\right)^{-n} (r^2 + R^2)^{\frac{n}{2}}}{
 D (2 + n) + C (r^2 + R^2)^{1 +\frac{ n}{2}}}\nonumber\\&&-\frac{1 - \left(1 + \frac{r^2}{R^2}\right)^{-n}}{r^2},\label{pr1}
 \end{eqnarray}
 \begin{eqnarray}
 \kappa p_t=\frac{\big(C (4 + n) (r^2 + R^2)^{1 + \frac{n}{2}}-D n (2 + n)\big)\left(1 + \frac{r^2}{R^2}\right)^{-n}}{
D (2 + n) (r^2 + R^2) + C (r^2 + R^2)^{2 + \frac{n}{2}}}.\label{pt1}\nonumber\\
\end{eqnarray}

\begin{figure}[htbp]
    \centering
        \includegraphics[scale=.6]{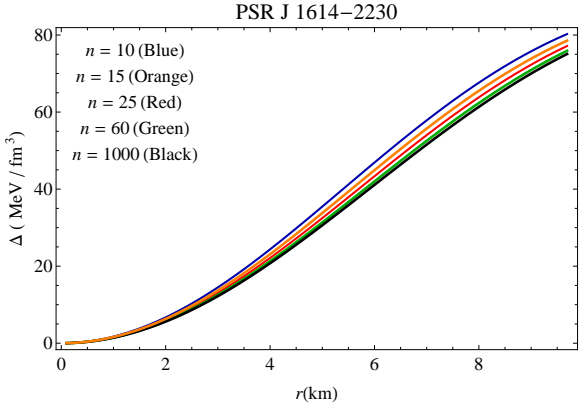}
       \caption{Anisotropic factor $\Delta$ is plotted against $r$ inside the stellar interior for the compact star PSR J1614-2230 for different values of `n' mentioned in the figure.\label{delta}}
\end{figure}

\begin{table*}[htbp]
\centering
\caption{The values of the constants $a,\,C$ and $D$ for the compact star PSR J1614-2230 whose observed mass and radius is given by $(1.97 \pm 0.04)M_{\odot}$ and $9.69_{-0.2}^{+0.2}$ km \cite{demo58}.}
{\begin{tabular}{@{}cccccccccc@{}} \hline
Objects & Estimated &Estimated & $n$ & $R$& $C$  &$D$ \\
&mass ($M_{\odot}$)&radius& (km) &(km$^{-2}$)\\
\hline
PSR J1614-2230 &$1.97$&$9.69$& 10& $31.2927$ & $3.5469\times10^{-18}$ & $0.1812$\\
&&& $15$ & $38.6228$ & $5.0306\times 10^{-27}$& $0.1618$                                       \\
&&& $20$ & $44.7698$& $3.0519\times 10^{-36}$&$0.1512$ \\
&&& $25$ & $50.1697$ & $9.8459\times 10^{-46}$& $0.1445$ \\
&&& $50$ & $71.2777$ & $7.1879\times 10^{-96}$ & $0.1303$ \\
&&& $60$ & $78.1406$ & $8.5446\times 10^{-117}$ & $0.1278$ \\
&&& $100$ & $101.033$ & $1.1423\times 10^{-203}$ & $0.1228$\\
&&& $500$ & $226.332$ & $1.3527\times 10^{-1180}$ & $0.1165$ \\
&&& $1000$ & $320.155$ & $1.3922\times 10^{-2508}$ & $0.1158$  \\
\hline
\end{tabular} \label{tab1}}
\end{table*}

\begin{table*}[ph]
\centering
\caption{The values of the constants $a,\,C$ and $D$ for the compact star EXO 1785-248 whose observed mass and radius is given by $(1.3\pm 0.2)M_{\odot}$ and $8.849_{-0.4}^{+0.4}$ km respectively \cite{ozel58}.}
{\begin{tabular}{@{}cccccccccc@{}} \hline
Objects & Estimated &Estimated & $n$ && $R$& $C$  &$D$ \\
&Mass ($M_{\odot}$)& Radius &&& (km) &(km$^{-2}$)\\
\hline
EXO 1785-248   & $1.4$&$8.8$ & 10&& $34.4082$ & $1.3027\times10^{-18}$ & $0.2912$\\
&&&$15$ && $42.3663$& $1.1916\times 10^{-27}$& $0.2703$                                       \\
&&& $20$ && $49.0506$& $4.6622\times 10^{-37}$& $0.2589$ \\
&&& $25$ && $54.9275$ & $9.6988\times 10^{-47}$& $0.2517$ \\
&&& $50$ && $77.9265$ & $7.8923\times 10^{-98}$ & $0.2365$ \\
&&& $60$ && $85.4094$ & $3.9005\times 10^{-119}$ & $0.2338$ \\
&&& $100$ && $110.38$ & $1.5579\times 10^{-207}$ & $0.2284$\\
&&& $500$ && $247.13$ & $1.0446\times 10^{-1199}$ & $0.2217$ \\
&&& $1000$ && $349.55$ & $9.3898\times 10^{-2547}$ & $0.2209$  \\
\hline
\end{tabular} \label{tab2}}
\end{table*}

\section{Boundary Condition}\label{bou}
To fix the different constants, in this section, we match our interior spacetime continuously to the exterior Schwarzschild spacetime,
\begin{eqnarray}
ds_+^{2}&=&\left(1-\frac{2M}{r}\right)^{-1}dr^{2}+r^{2}\left(d\theta^{2}+\sin^{2}\theta d\phi^{2}\right)\nonumber\\&&-\left(1-\frac{2M}{r}\right)dt^{2},
\end{eqnarray}
outside the event horizon, i.e., $r>2M$, where $M$ is a constant representing the total mass of the compact star corresponding to our interior spacetime:
 \begin{eqnarray}
ds_{-}^2 & = &\left(1 + \frac{r^2}{R^2}\right)^{n} dr^2+r^2(d\theta^2+\sin^2 \theta d\phi^2) - \nonumber\\&&\left(D + \frac{C (r^2 + R^2)^{1 + \frac{n}{2}}}{2 + n}\right)^2 dt^2,
\end{eqnarray}
A smooth matching of the metric potentials across the boundary is given by the first fundamental
form, i.e., at the boundary $r=r_b$,
\begin{eqnarray}\label{b1}
g_{rr}^+=g_{rr}^-,~g_{tt}^+=g_{tt}^-,
\end{eqnarray}
and the second fundamental form implies
\begin{eqnarray}\label{b2}
p_i(r=r_b-0)=p_i(r=r_b+0).
\end{eqnarray}
where $i$ takes the value $r$ and $t$.\\
From the boundary conditions $g_{tt}^+=g_{tt}^-$ and $p_r(r_b)=0$, we get the following two equations:
\begin{eqnarray}
1-\frac{2M}{r_b}&=&\left(D + \frac{C (r_b^2 + R^2)^{1 + \frac{n}{2}}}{2 + n}\right)^2,\label{o1}\\
\frac{1 - \left(1 + \frac{r_b^2}{R^2}\right)^{-n}}{r_b^2}&=&\frac{
 2 C (2 + n) \left(1 + \frac{r_b^2}{R^2}\right)^{-n} (r_b^2 + R^2)^{\frac{n}{2}}}{
 D (2 + n) + C (r_b^2 + R^2)^{1 +\frac{ n}{2}}}\label{o2}\nonumber\\
\end{eqnarray}
\begin{itemize}
  \item {\bf Determination of $R$:} ~Now the boundary condition $g_{rr}^+=g_{rr}^-$, implies,
\begin{eqnarray}
R &=& \frac{r_b}{\sqrt{\left(1 -\frac{2M}{r_b}\right)^{-\frac{1}{n}} - 1}}\label{b1}
\end{eqnarray}
  \item {\bf Determination of $C$ and $D$ :}~ Solving the equations (\ref{o1}) and (\ref{o2}) we obtain the expression $C$ and $D$ as,
  \begin{eqnarray}
 C&=&-\frac{(R^2 + r_b^2)^{-\frac{n}
   {2}} \sqrt{\frac{-2 M + r_b}{r_b}}\left(1-\left(1 + \frac{r_b^2}{R^2}\right)^n\right)}{
 2 r_b^2} \label{b2}\nonumber\\
 \\
 D&=&-\frac{C(r_b^2 + R^2)^{1 + \frac{n}{2}}}{2 + n} + \sqrt{1 - \frac{2M}{r_b}};\label{b3}
 \end{eqnarray}
 \end{itemize}
So, we have matched our interior spacetime to the exterior Schwarzschild spacetime at the boundary $r=r_{b}$. Obviously it is clear that the metric coefficients are continuous at $r=r_b$, but the transverse pressure $p_t$ does not vanish at the boundary and therefore it is not continuous at the junction surface. To take care of this situation, let us use the Darmois-\cite{i1,i2} formation to determine the surface stresses at the junction boundary. The intrinsic surface stress energy tensor $S_{ij}$ is given by Lancozs equations in the following form
\begin{equation}
S^{i}_{j}=-\frac{1}{8\pi}(\kappa^{i}_j-\delta^{i}_j\kappa^{k}_k),
\end{equation}
where $\kappa_{ij}$ represents the discontinuity in the second fundamental form is written as,
\begin{equation}
\kappa_{ij}=K_{ij}^{+}-K_{ij}^{-},
\end{equation}
$K_{ij}$ being the second fundamental form is presented by
\begin{equation}
K_{ij}^{\pm}=-n_{\nu}^{\pm}\left[\frac{\partial^{2}X_{\nu}}{\partial \xi^{i}\partial\xi^{j}}+\Gamma_{\alpha\beta}^{\nu}\frac{\partial X^{\alpha}}{\partial \xi^{i}}\frac{\partial X^{\beta}}{\partial \xi^{j}} \right]|_S,
\end{equation}
where $n_{\nu}^{\pm}$ are the unit normal vectors defined by,
\begin{equation}
n_{\nu}^{\pm}=\pm\left|g^{\alpha\beta}\frac{\partial f}{\partial X^{\alpha}}\frac{\partial f}{\partial X^{\beta}}  \right|^{-\frac{1}{2}}\frac{\partial f}{\partial X^{\nu}},
\end{equation}
with $n^{\nu}n_{\nu}=1$. Where $\xi^{i}$ is the intrinsic coordinate on the shell. $-$ and $+$ corresponds to interior and exterior Schwarzschild spacetime respectively.\par

\begin{figure}[htbp]
    \centering
        \includegraphics[scale=.6]{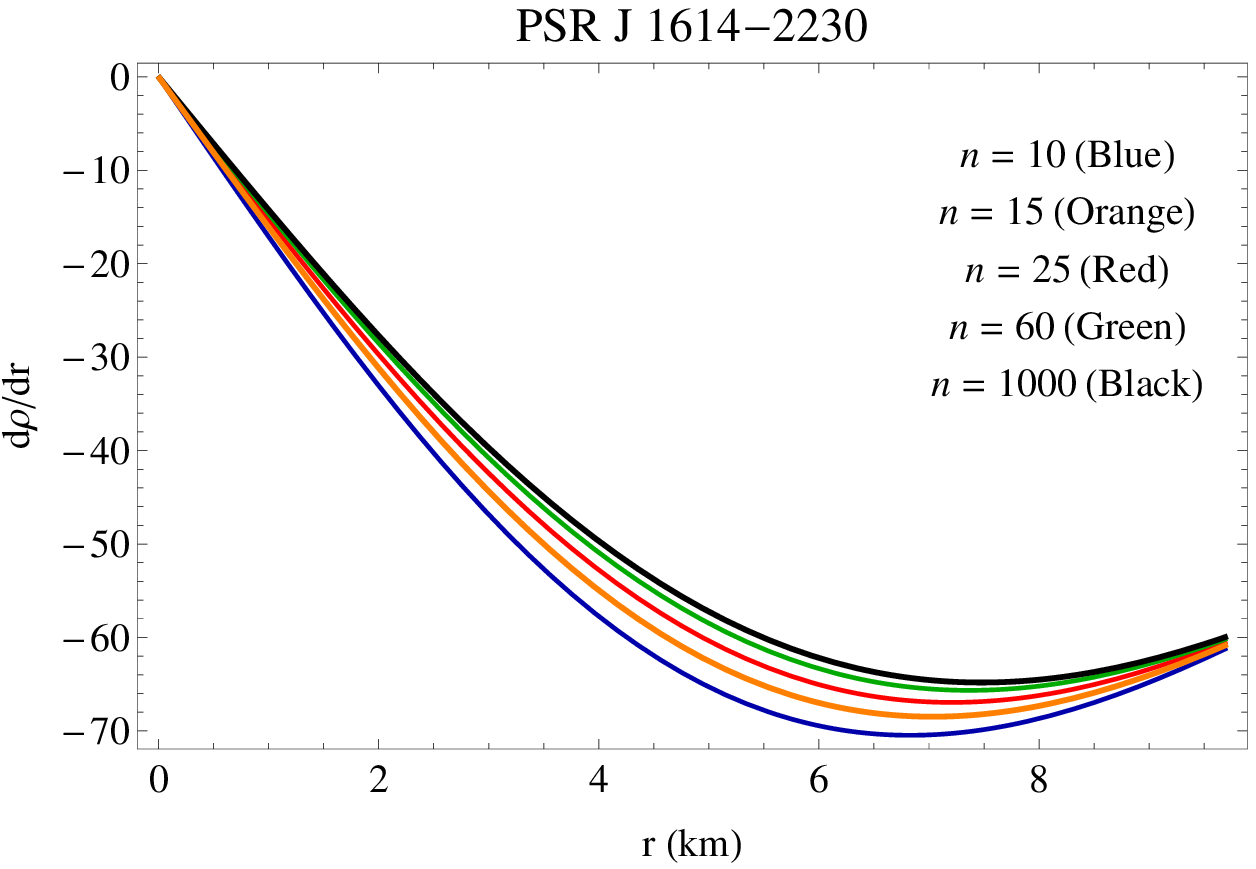}
        \includegraphics[scale=.6]{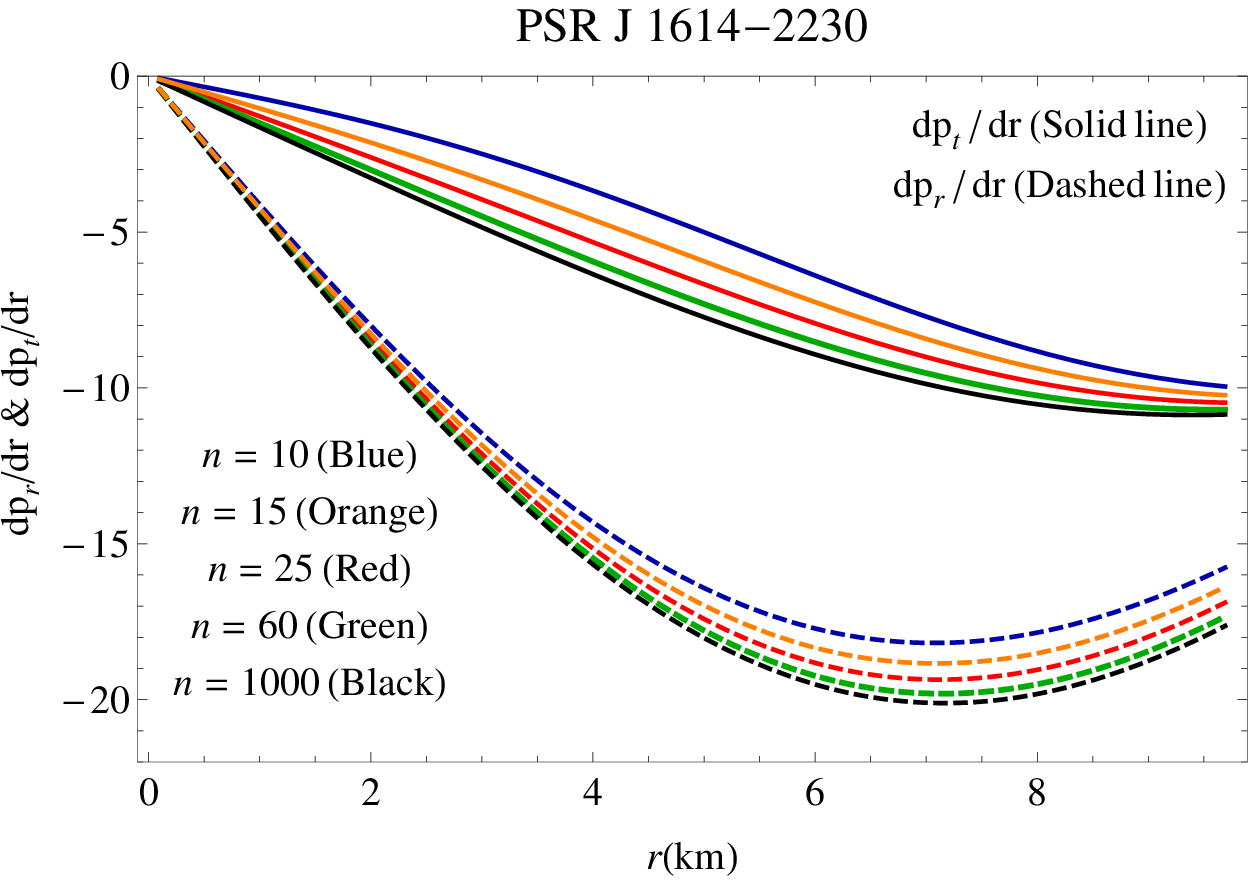}
       \caption{(Top) $\frac{d\rho}{dr}$  and (bottom) $\frac{dp_r}{dr},\,\frac{dp_t}{dr}$ are plotted against $r$ inside the stellar interior for the compact star PSR J1614-2230 for different values of `n' mentioned in the figures.}\label{drho}
\end{figure}

Using the spherical symmetry nature of the spacetime surface stress energy tensor can be written as $S^{i}_j=diag(-\sigma,\mathcal{P})$. Where $\sigma$ and $\mathcal{P}$ being the surface energy density and surface pressure respectively.\\
The expression for surface energy density $\sigma$ and the surface pressure $\mathcal{P}$ at the junction surface $r = r_b$ are obtained as,
\begin{eqnarray}
\sigma &=&-\frac{1}{4\pi r_b}\left[\sqrt{1-\frac{2M}{r_b}}-\left(1 + \frac{r_b^2}{R^2}\right)^{-\frac{n}{2}}\right],\label{sig}\\
\mathcal{P}&=&\frac{1}{8\pi r_b}\bigg[\frac{1-\frac{M}{r_b}}{\sqrt{1-\frac{2M}{r_b}}}-\frac{\left(1 + \frac{r_b^2}{R^2}\right)^{-\frac{n}{2}}}{D (2 + n) + C (r_b^2 + R^2)^{1 + \frac{n}{2}}}\times \nonumber\\&&\Big\{D (2 + n) + C (r_b^2 + R^2)^{\frac{n}{2}}\big((3 + n) r_b^2 + R^2\big)\Big\}\bigg].\nonumber\\
\end{eqnarray}

\section{Physical analysis of the present model}\label{pa}
\begin{figure}[htbp]
    \centering
        \includegraphics[scale=.6]{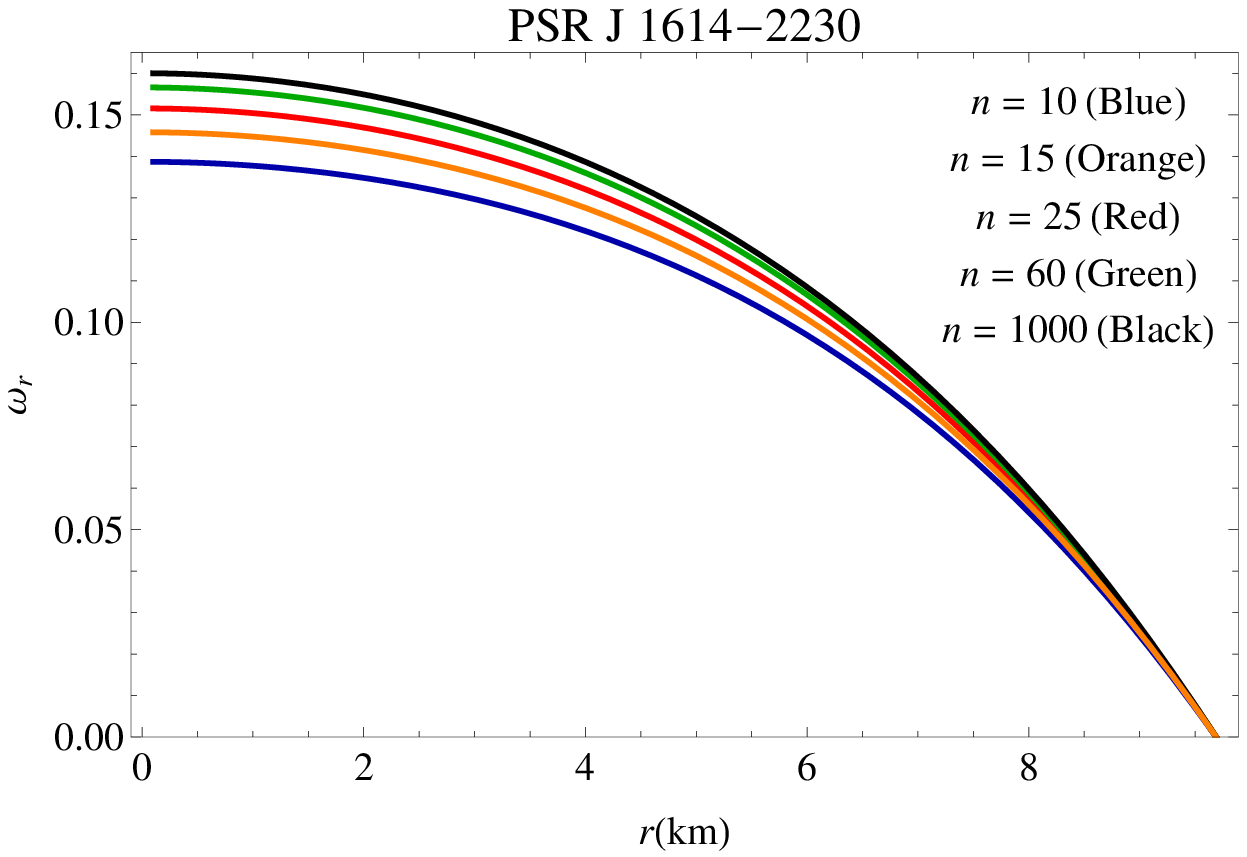}
         \includegraphics[scale=.6]{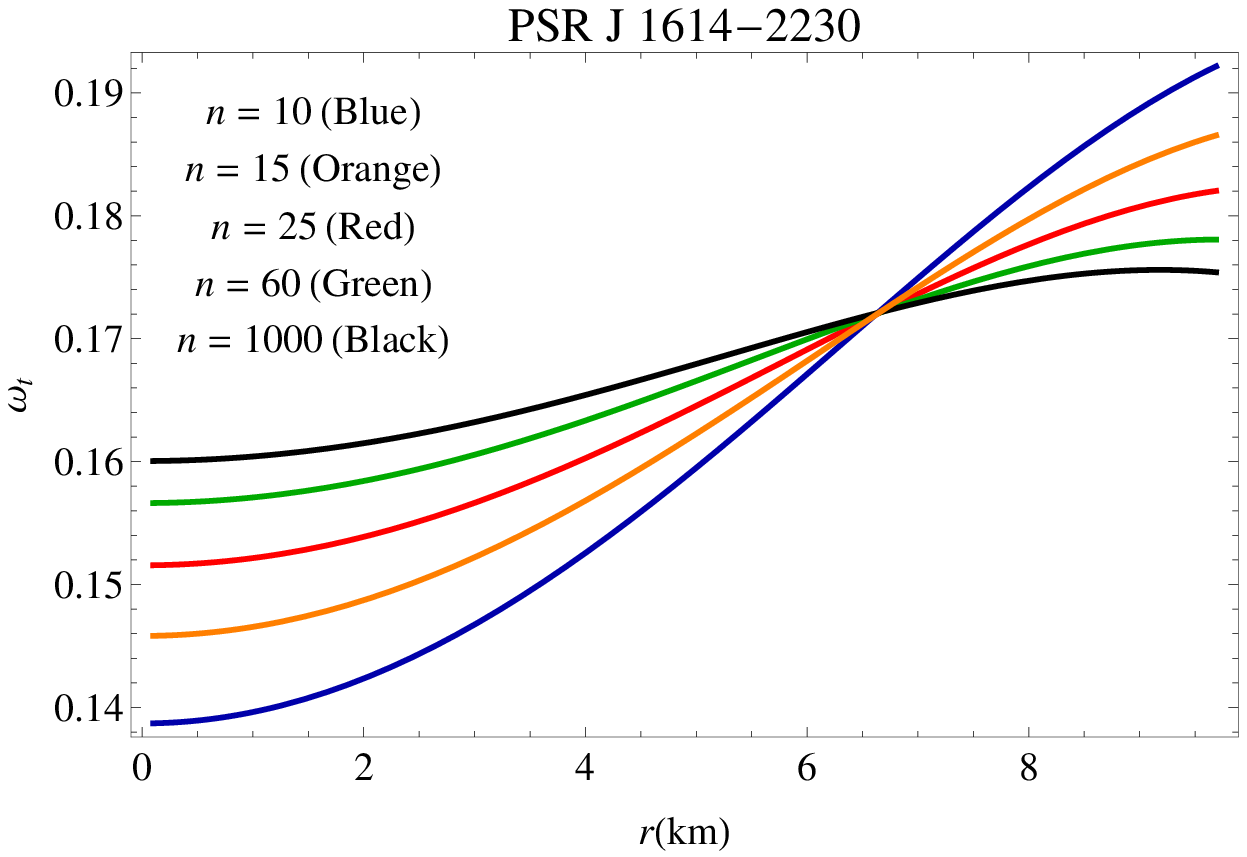}
           \includegraphics[scale=.6]{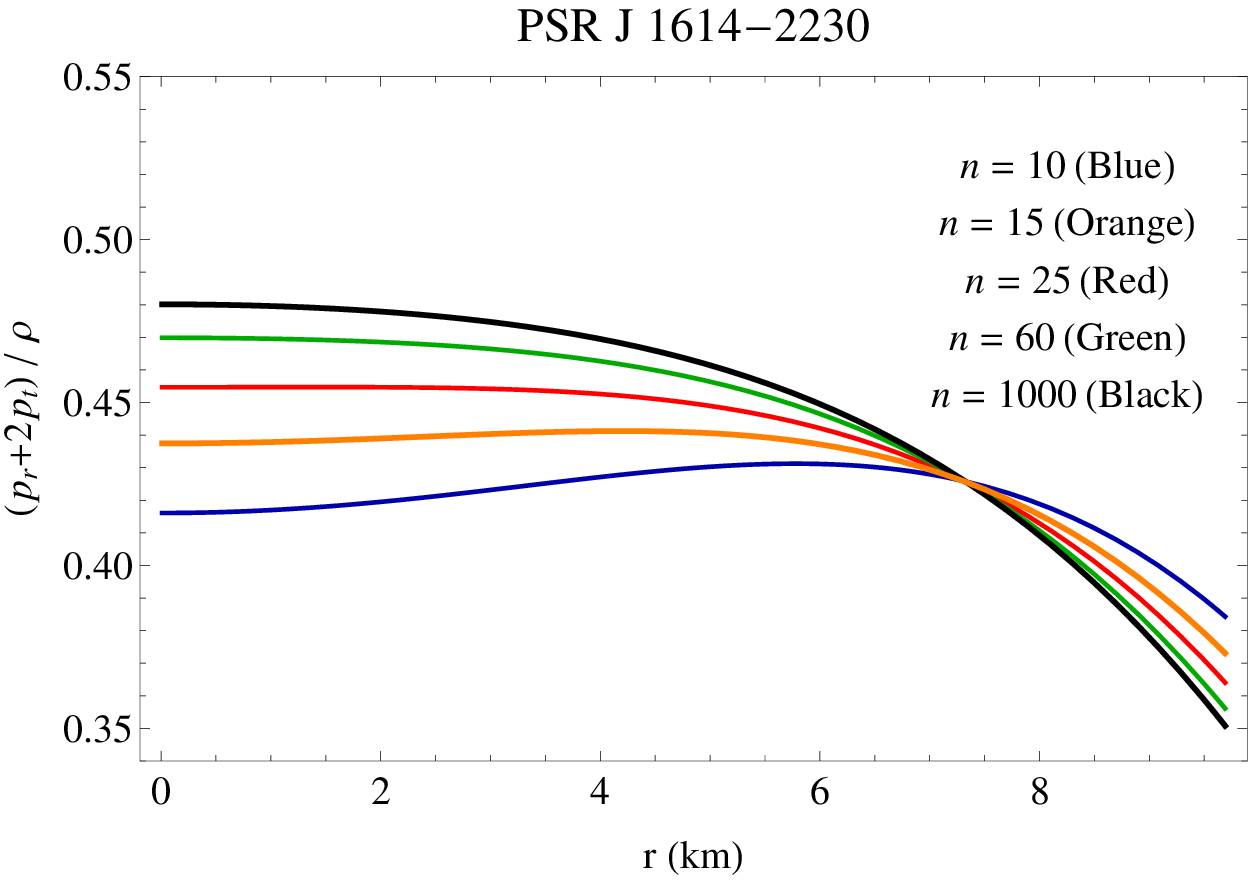}
       \caption{ Equation of state parameters (top) $\omega_r$, (middle) $\omega_t$ and (bottom) $\frac{p_r+2p_t}{\rho}$ are plotted against $r$ inside the stellar interior for the compact star PSR J1614-2230 for different values of `n' mentioned in the figures.}\label{omega}
\end{figure}

\begin{enumerate}
  \item {\bf Regularity of the metric functions at the center:} ~We observe from eqns.~(\ref{enu1}) and (\ref{elambda1}) that
the metric potentials take the following values at the center of the star:
\begin{eqnarray} B^2|_{r=0}&=&1,\\A^2|_{r=0}&=&\Big(\frac{D (2 + n) + C R^{2+n}}{2 + n}\Big)^2,
 \end{eqnarray}
The above two equations imply that metric functions are free from singularity and positive at the center.
\item {\bf Behavior of pressure and density:}~ From eqns. (\ref{rho1}), (\ref{pr1}) and (\ref{pt1}) the central density and central pressure are obtained as,
\begin{eqnarray}
\rho_c&=&\frac{3 n}{\kappa R^2}~>0,\label{rho2}\\
p_c&=&\frac{-D n (2 + n) + C (4 + n)R^{2+n}}{R^2 \big(D (2 + n) + C R^{2+n}\big)}>0.\label{pr2}
\end{eqnarray}
From eqn. (\ref{rho2}), we get, $n>0$. Therefore we can conclude that the dimensionless parameter $n$ can never be negative. So $n \in R^{+}$, $R$ being the set of real numbers.\\
Now eqn. (\ref{pr2}) holds if,
$$\frac{C}{D}>\frac{n(2+n)}{(4+n)R^{2+n}},$$
Also it is well known that for a physically acceptable model, $\rho-p_r-2p_t$ should be positive everywhere within the stellar interior. By employing $\rho-p_r-2p_t$ at the center of the star we get,
$$\frac{n(2+n)}{2R^{n+2}}~>\frac{C}{D},$$
Hence we obtain a reasonable bound for $\frac{C}{D}$ as,
$$\frac{n(2+n)}{(4+n)R^{2+n}}<~\frac{C}{D}~<\frac{n(2+n)}{2R^{n+2}}.$$
The surface density of the star is obtained as,
\begin{eqnarray}
\kappa \rho_s&=&\frac{1 - \left(1 + \frac{r_b^2}{R^2}\right)^{-n}}{r_b^2} + \frac{2 n \left(1 + \frac{r_b^2}{R^2}\right)^{-1 - n}}{R^2}.
   \end{eqnarray}
   The equation of state parameters $\omega_r$ and $\omega_t$ are obtained from the following formulae,
   $$\omega_r=\frac{p_r}{\rho},~\omega_t=\frac{p_t}{\rho}.$$
   The profiles of $\omega_r$ and $\omega_t$ have been plotted in Fig.~\ref{omega}.\par

 Next we are interested to find out the pressure and density gradient denoted as follows :\\
 (i)~Density gradient : $\frac{d\rho}{dr}$, (ii)~Radial pressure gradient : $\frac{dp_r}{dr}$, (iii)~Transverse pressure gradient : $\frac{dp_t}{dr}$.\\
Differentiating the eqns. (\ref{rho1}), (\ref{pr1}) and (\ref{pt1}) with respect to `r' we get the following expressions for $\rho',\, p_r'$ and $p_t'$ (overdashed denotes differentiation with respect to `r') as,
\begin{eqnarray}
\kappa \rho'&=&\frac{2}{r^3}\Big[\frac{\left(1 + \frac{r^2}{R^2}\right)^{-n}}{(r^2 + R^2)^2}\Big\{(1-n-2 n^2)r^4\nonumber\\&&+ (2 + n) r^2 R^2 +
    R^4\Big\}-1\Big], \label{rhod}
    \\
    \kappa p_r'&=&\frac{2 - 2 \left(1 +\frac{r^2}{R^2}\right)^{-n} }{r^3} - \frac{2 n  \left(1 +\frac{r^2}{R^2}\right)^{-1-n}}{
 r R^2}\nonumber\\&&-
 2 C (2 + n) r \Psi_1(r)\\
 \kappa p_t' &=&\frac{2 r \left(1 + \frac{r^2}{R^2}\right)^{-n}}{\left(D (2 + n) (r^2 + R^2) + C (r^2 + R^2)^{2 + \frac{n}{2}}\right)^2}\times\nonumber\\&&\Big\{D^2 n (1 + n) (2 + n)^2 + C D n^2 (2 + n)\times \nonumber\\&&(r^2 + R^2)^{1 + \frac{n}{2}} -
 C^2 (1 + n) (4 + n) \times \nonumber\\&&(r^2 + R^2)^{2 + n}\Big\},\label{ptd}
\end{eqnarray}
where,
\begin{eqnarray*}
\Psi_1(r)&=&\frac{D n (2 + n) + 2 C (1 + n) (r^2 + R^2)^{1 + \frac{n}{2}}}{\Big(D (2 + n) + C (r^2 + R^2)^{1 + \frac{n}{2}}\Big)^2}\Omega,\\
\Omega&=&\frac{(r^2 + R^2)^{-1 + \frac{n}{2}}}{\left(1 +\frac{r^2}{R^2}\right)^{n}}.
\end{eqnarray*}
All of $\rho',\,p_r',\,p_t'$ should be negative for $r~\in~(0,\,r_b)$. We shall verify it with the help of graphical analysis.
\end{enumerate}

\begin{table*}[ph]
\centering
\caption{The numerical values of the central density $\rho_c$, surface density $\rho_s$, central pressure $p_c$, surface transverse pressure and the value of the radial adiabatic index $\Gamma_r$ at the center have been obtained for different values of `n' for the compact star PSR J1614-2230.}
{\begin{tabular}{@{}cccccccccc@{}} \hline
n&& $\rho_c$ && $\rho_s$ & $p_c$ & $\Gamma_{r0}$ & $p_t(r_b)$\\
&& gm.cm$^{-3}$&& gm.cm$^{-3}$& dyne.cm$^{-2}$&& dyne.cm$^{-2}$\\
\hline
 $10$ && $1.6448\times10^{15}$ && $7.4341\times10^{14}$ & $2.0506\times10^{35}$ &$1.9691$  &$1.28407\times 10^{35}$       \\
 $15$ && $1.6196\times10^{15}$ && $7.4950\times10^{14}$ & $2.1228\times10^{35}$   &$2.06909$   &$1.25671\times 10^{35}$                             \\
 $20$ && $1.6071\times10^{15}$  && $7.5258\times10^{14}$ & $2.1583\times10^{35}$   &$2.12086$   &$1.24282\times 10^{35}$                     \\
 $25$ &&  $1.5998\times10^{15}$  && $7.5445\times10^{14}$  & $2.1795\times10^{35}$ &$2.15254$   &$1.23442\times 10^{35}$                                           \\
 $50$ &&  $1.5851\times10^{15}$ && $7.5822\times10^{14}$ & $2.2213\times10^{35}$  &$2.21738$    &$1.21746\times 10^{35}$                                    \\
 $60$ &&  $1.5827\times10^{15}$ && $7.5885\times10^{14}$ & $2.2282\times10^{35}$   &$2.22839$   &$1.21461\times 10^{35}$                           \\
 $100$ && $1.5779\times10^{15}$ && $7.6012\times10^{14}$& $2.2420\times10^{35}$ &$2.25058$       &$1.2089\times 10^{35}$                     \\
 $500$ && $1.5721\times10^{15}$ && $7.6165\times10^{14}$& $2.2585\times10^{35}$ &$2.27754$       &$1.20202\times 10^{35}$                 \\
 $1000$ && $1.5714\times10^{15}$ && $7.6184\times10^{14}$ & $2.2606\times10^{35}$ &$2.28094$     &$1.20116\times 10^{35}$                             \\
\hline
\end{tabular} \label{tab3}}
\end{table*}

\begin{table*}[ph]
\centering
\caption{The numerical values of the central density $\rho_c$, surface density $\rho_s$, central pressure $p_c$, surface transverse pressure and the value of the radial adiabatic index $\Gamma_r$ at the center have been obtained for different values of `n' for the compact star EXO 1785-248.}
{\begin{tabular}{@{}cccccccccc@{}} \hline
n&& $\rho_c$ && $\rho_s$ & $p_c$ & $\Gamma_{r0}$ & $p_t(r_b)$\\
&& gm.cm$^{-3}$&& gm.cm$^{-3}$& dyne.cm$^{-2}$&& dyne.cm$^{-2}$\\
\hline
 $10$ && $1.3604\times10^{15}$ && $6.9696\times10^{14}$ & $8.8028\times10^{34}$ &$2.3168$  &$8.5476\times 10^{34}$       \\
 $15$ && $1.3460\times10^{15}$ && $7.0253\times10^{14}$ & $9.2288\times10^{34}$   &$2.4278$   &$8.2944\times 10^{34}$                             \\
 $20$ && $1.3389\times10^{15}$  && $7.0537\times10^{14}$ & $9.4396\times10^{34}$   &$2.4849$   &$8.1657\times 10^{34}$                     \\
 $25$ &&  $1.3346\times10^{15}$  && $7.0708\times10^{14}$  & $9.5653\times10^{34}$ &$2.5196$   &$8.0878\times 10^{34}$                                           \\
 $50$ &&  $1.3262\times10^{15}$ && $7.1055\times10^{14}$ & $9.8151\times10^{34}$  &$2.5906$    &$7.9305\times 10^{34}$                                    \\
 $60$ &&  $1.3248\times10^{15}$ && $7.1113\times10^{14}$ & $9.8565\times10^{34}$   &$2.6026$   &$7.9041\times 10^{34}$                           \\
 $100$ && $1.3220\times10^{15}$ && $7.1230\times10^{14}$& $9.9392\times10^{34}$ &$2.6268$       &$7.8511\times 10^{34}$                     \\
 $500$ && $1.3186\times10^{15}$ && $7.1370\times10^{14}$& $1.0038\times10^{35}$ &$2.6561$       &$7.7872\times 10^{34}$                 \\
 $1000$ && $1.3182\times10^{15}$ && $7.1388\times10^{14}$ & $1.0050\times10^{35}$ &$2.6598$     &$7.7792\times 10^{34}$                             \\
\hline
\end{tabular} \label{tab4}}
\end{table*}

\subsection{The compact star PSR J1614-2230 and EXO 1785-248}

To obtain the values of the constants $R,\,C$ and $D$ in the expressions of different model parameters, we have considered two compact stars PSR J1614-2230 and EXO 1785-248. The observed and estimated masses and radii of these two stars have been given in table~\ref{tab1} and table~\ref{tab2} respectively. Using the expressions for $R,\,C$ and $D$ from eqns. (\ref{b1})-(\ref{b3}), the numerical values of these parameters have been determined for different values of $n$ and presented in table~\ref{tab1} and table~\ref{tab2}.\\
By using the mentioned values in table~\ref{tab1} for the compact star star PSR J1614-2230, we generated the plots of matter density $(\rho)$, radial pressure ($p_r$), transverse pressure ($p_t$), anisotropic factor $(\Delta)$ in Figs.~\ref{rho}, \ref{pr} and \ref{delta} respectively. The matter density $\rho$ is positive, finite and monotonically decreasing function of `r', i.e., the maximum values of these physical model parameters are attained at the center of the star. The radial and transverse pressure $p_r,\,p_t$ show the same behavior as $\rho$. Fig.~\ref{delta} shows that $\Delta>0$ for our model, i.e., anisotropic force is repulsive in nature and it is necessary for the construction of compact object \cite{gm}. Moreover at the center of the star $\Delta$ vanishes. $\rho',\,p_r',\,p_t'$ all are plotted in Fig.~\ref{drho} and the figures indicate that all of them take negative value and it once again verifies that $\rho,\,p_r,\,p_t$ are monotonic decreasing. The equation of state parameters $\omega_r$ and $\omega_t$ are plotted in Fig. \ref{omega}. From the profiles, we see that $\omega_r$ is monotonic decreasing and $\omega_t$ is monotonic increasing function of $r$ also $0<\omega_r,\,\omega_t<1$, indicate that the underlying matter distribution is non-exotic in nature. The ratio of stress tensor to energy $\frac{p_r+2p_t}{\rho}$ is monotonic decreasing and has been shown graphically in Fig. \ref{omega}.\\
One interesting thing we can note that, for larger values of `n', for the model parameters $\omega_t$ and $\frac{p_r+2p_t}{\rho}$, the central values of these two physical quantities increase and at $r=7.3$ km., irrespective of `n', all the curves coincide. Moreover for $ 0 < r <7.3 $, as `n' increases, the profile corresponding to small value of `n' is dominated by the profile corresponding to the value of `n' larger than previous one. On the other hand for $ 7.3 < r < r_b $, the nature of the curves for these two physical quantities become the reverse of the former.

\subsection{ Energy Conditions }
To be a physically reasonable model, one of the most important properties that should be satisfy by our model is that the energy
conditions {\em viz}, null energy condition (NEC), weak energy
condition (WEC), strong energy condition (SEC) and dominant energy condition (DEC). These energy conditions are obeyed if the following inequalities hold simultaneously:
\begin{eqnarray}\label{1}
NEC:&:& T_{\mu \nu}\beta^\mu \beta^\nu \ge 0~\mbox{or}~ \rho+p_i \geq  0, \label{3}\\
WEC: &:& T_{\mu \nu}\alpha^\mu \alpha^\nu \ge 0~\mbox{or}~\rho \geq  0,~\rho+p_i \ge 0, \label{2k}\\
DEC:&:& T_{\mu \nu}\alpha^\mu \alpha^\nu \ge 0 ~\mbox{or}~ \rho \ge |p_i|,  \\ \label{4}
SEC: &:& T_{\mu \nu}\alpha^\mu \alpha^\nu - {1 \over 2} T^\lambda_\lambda \alpha^\sigma \alpha_\sigma \ge 0 ~\mbox{or}~ \rho+\sum_i p_i \ge 0.\nonumber\\\label{4k}
\end{eqnarray}

\begin{figure}[htbp]
    \centering
        \includegraphics[scale=.52]{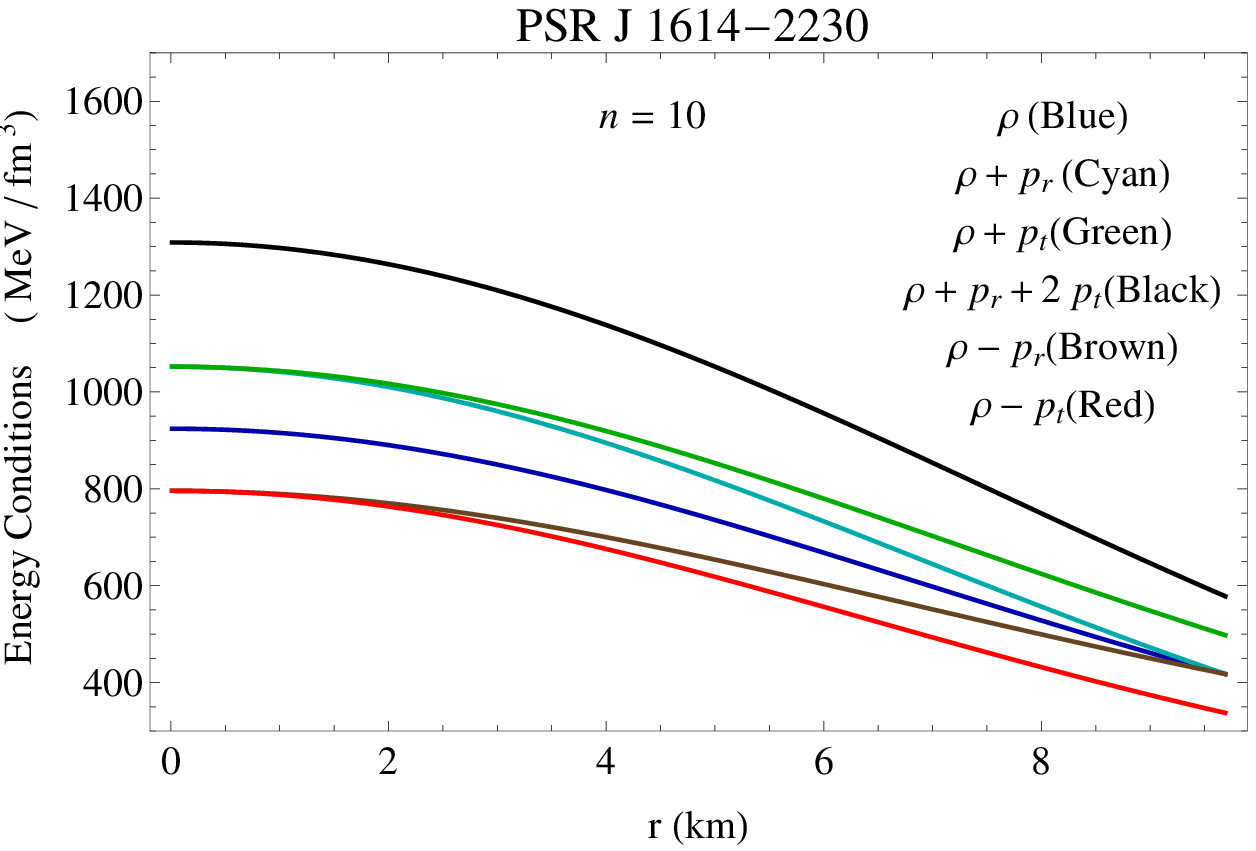}
        \includegraphics[scale=.52]{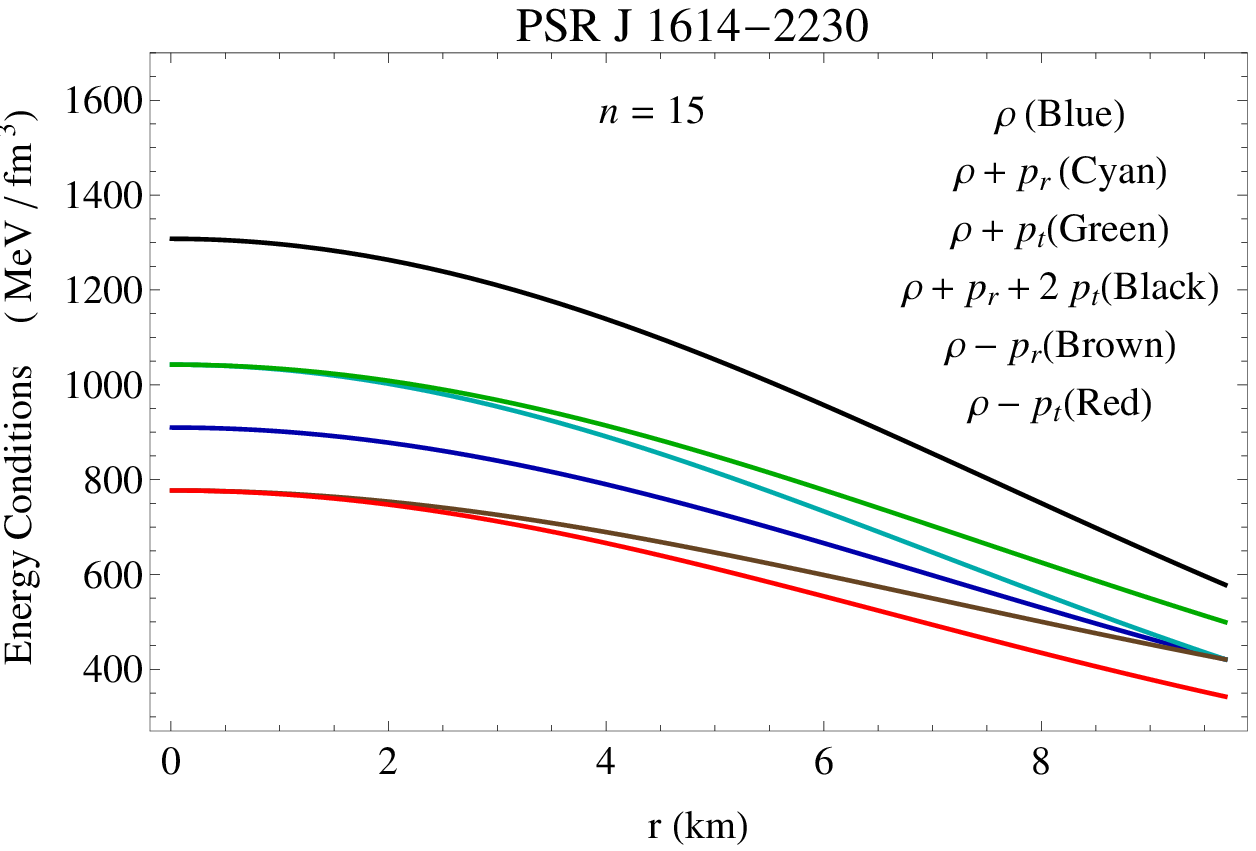}
        \includegraphics[scale=.52]{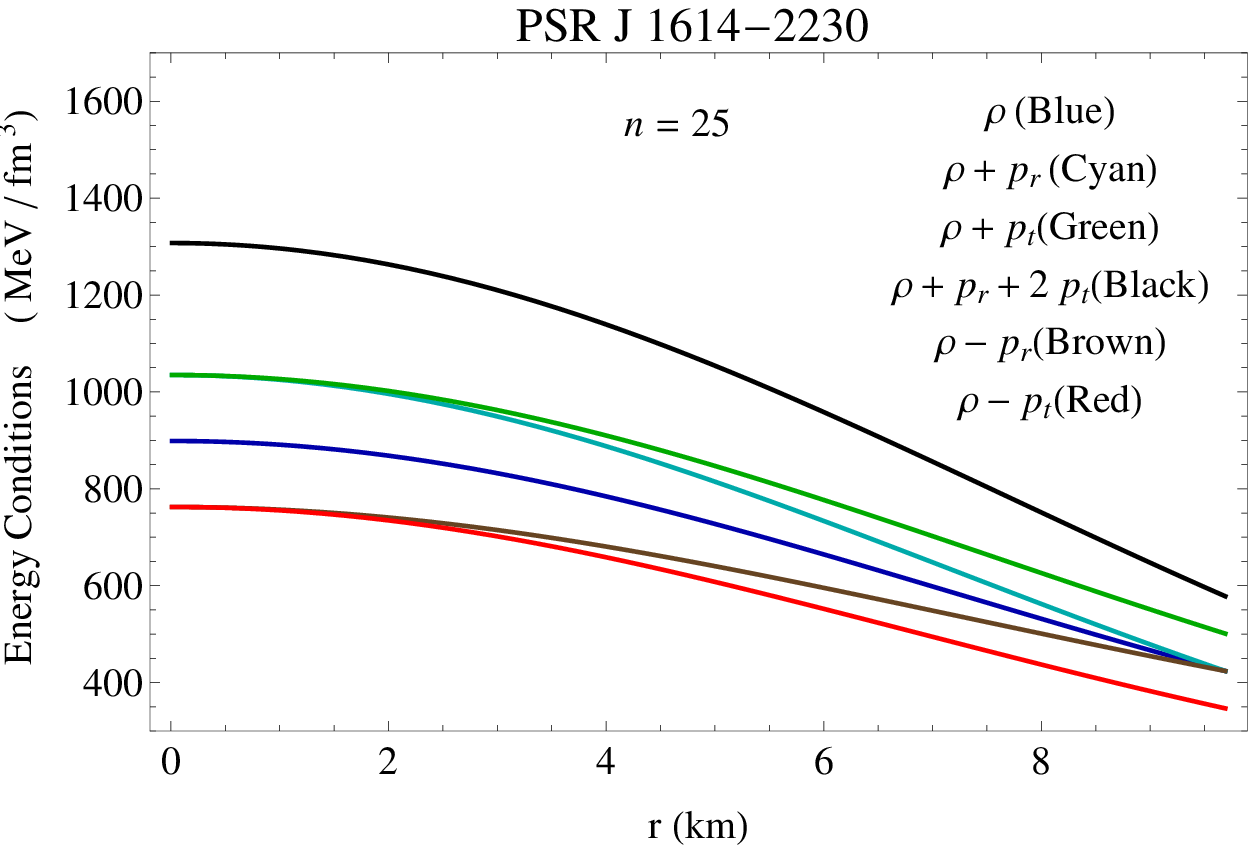}
        \includegraphics[scale=.52]{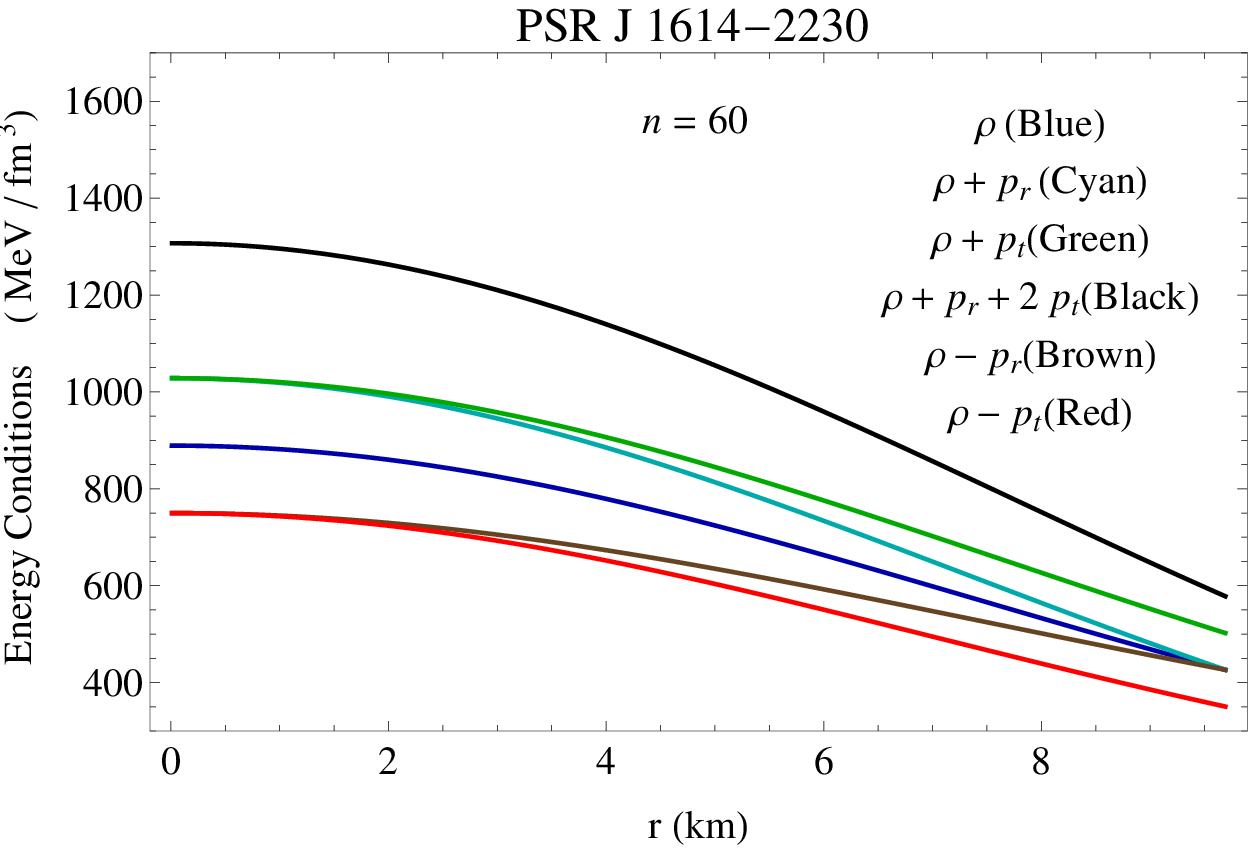}
        \includegraphics[scale=.52]{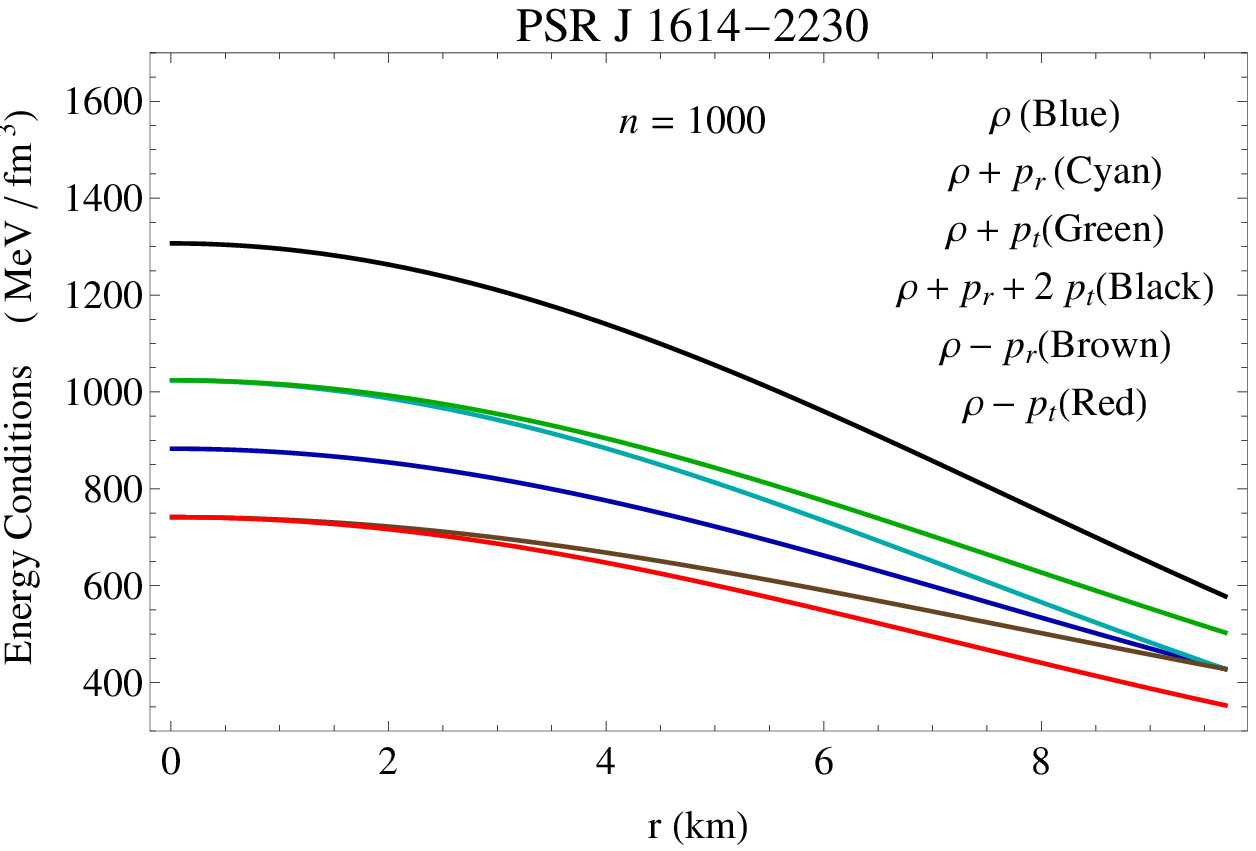}
       \caption{All the energy conditions are plotted against $r$ inside the stellar interior for the compact star PSR J1614-2230 for different values of `n' mentioned in the figures.\label{ec}}
\end{figure}

Where $i$ takes the value $r$ and $t$ for radial and transverse pressure. $~\alpha^\mu$ and $\beta^\mu$ are time-like vector and null vector respectively and $T^{\mu \nu}\alpha_\mu $ is non space-like vector. To check all the inequality stated above, we have drawn the profiles of left hand sides of (\ref{3})-(\ref{4k}) in Fig.~\ref{ec} in the interior of the compact star PSR J1614-2230 for different values of `n'. The figures show that all the energy conditions are satisfied by our model of compact star.

\subsection{ The behavior of mass function}
 The gravitational mass in a sphere of radius `r' is given by,
\begin{eqnarray}
m(r)&=&4\pi\int_0^{r}\omega^{2}\rho(\omega)d\omega=\frac{r}{2}\left[1 - \left(1 + \frac{r^2}{R^2}\right)^{-n}\right],\nonumber\\
\label{4}
\end{eqnarray}
The compactness factor $u(r)$ and surface redshift $z_s$ is defined by,
\begin{eqnarray}
u(r)&=&\frac{4\pi}{r}\int_0^{r}\omega^{2}\rho(\omega)d\omega=\frac{1}{2}\left[1 - \left(1 + \frac{r^2}{R^2}\right)^{-n}\right],\label{5}\nonumber\\
\\
z_s &=&\frac{1}{\sqrt{1-2u(r_b)}}-1.
\end{eqnarray}
One can easily check that $\lim_{r \rightarrow 0} m(r)=0$, which indicates that the mass function is regular at the center of the star. What is more, $\left(1+\frac{r^2}{R^2}\right)^n>1$, therefore, $\left(1+\frac{r^2}{R^2}\right)^{-n}<1$ and hence $m(r)>0$. As `n' increases, $\left(1+\frac{r^2}{R^2}\right)^{-n}$ decreases and therefore mass function increases. So, m(r) is a monotonic increasing function of $r$. The twice maximum allowable ratio of mass to the radius are obtained as, $0.5997$ and $0.4693$ for the compact stars PSR J1614-2230 and EXO 1785-248 respectively, and these values lie in the range $\frac{2M}{r_b}<\frac{8}{9}$, i.e., the Buchdahl's limit is satisfied \cite{buch}. The surface redshift of these two stars are obtained as $0.5806$ and $0.372699$ respectively.

\begin{figure}[htbp]
    \centering
        \includegraphics[width=6 cm]{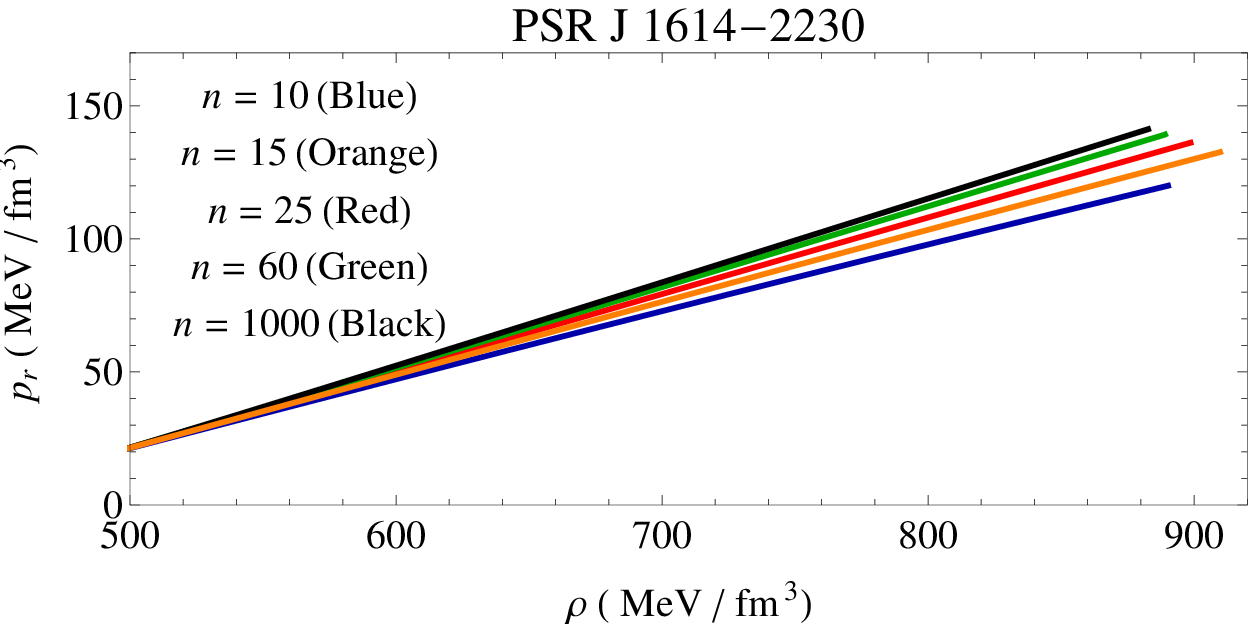}
         \includegraphics[width=6cm]{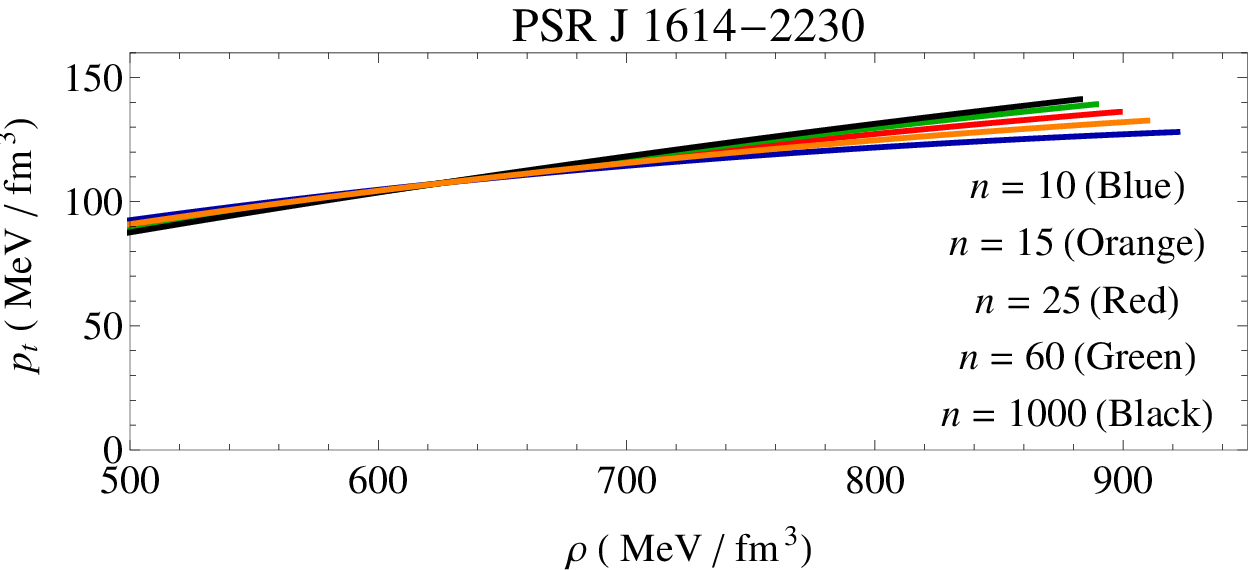}
       \caption{(Top) $p_r$ versus $\rho$ and (bottom) $p_t$ versus $\rho$ are plotted against $r$ inside the stellar interior for the compact star PSR J1614-2230 for different values of `n' mentioned in the figures. \label{eos}}
\end{figure}
\subsection{Equation of state }
To construct a model of a compact star, it is a very common choice among the researchers to use any specific equation of state (EOS) which gives a relationship between the pressure and density.
We still do not know the relationship between the matter density and the pressures. To check the variation of pressure with density, we have plotted $p_r$ versus $\rho$ and $p_t$ versus $\rho$ in Fig.~\ref{eos}, from which one can predict possible
EoS. For the complexity in the expressions
of $\rho,\,p_r$ and $p_t$, it is very difficult to obtain a well known
relation between them. With the help of the numerical analysis, we have obtained the best fitted curve which gives us a prediction about the relationship between the matter density and pressure. We have drawn the profiles for the compact star PSR J1614-2230 by taking different values of the dimensionless constant `n' mentioned in the figures.

\subsection{ Mass-radius curves}

\begin{figure}[htbp]
    \centering
        \includegraphics[scale=.62]{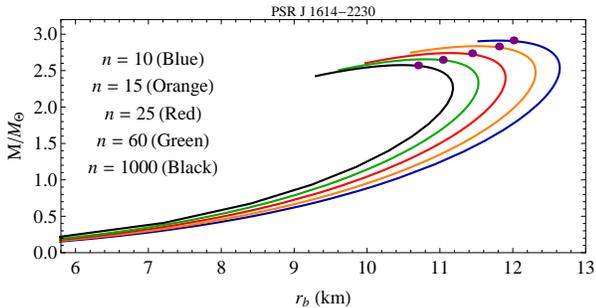}
       \caption{Maximum allowable mass versus radius is plotted against $r$ inside the stellar interior for different values of `n' mentioned in the figure.\label{mr}}
\end{figure}

The mass-radius relation for different values of the dimensionless parameter `n' for the compact star PSR J1614-2230 is shown in Fig.~\ref{mr}. In the following table we have given the maximum allowable mass for different values of `n' and we have also obtained the corresponding radius from the figure.
\begin{center}
\begin{tabular}{ c | c |c  }
n &   Maximum mass  & Radius \\
&($M/M_{\odot}$)& (km)\\
\hline
10 & 2.915 & 12.01 \\
15 & 2.831 & 11.81\\
25 & 2.738 & 11.44 \\
60 & 2.648 & 11.04 \\
1000&2.572& 10.7      \\
\hline
\end{tabular}
\end{center}

It is evident from the calculation that the maximum mass of compact star decreases as `n' increases.


\begin{figure}[htbp]
    \centering
        \includegraphics[scale=.6]{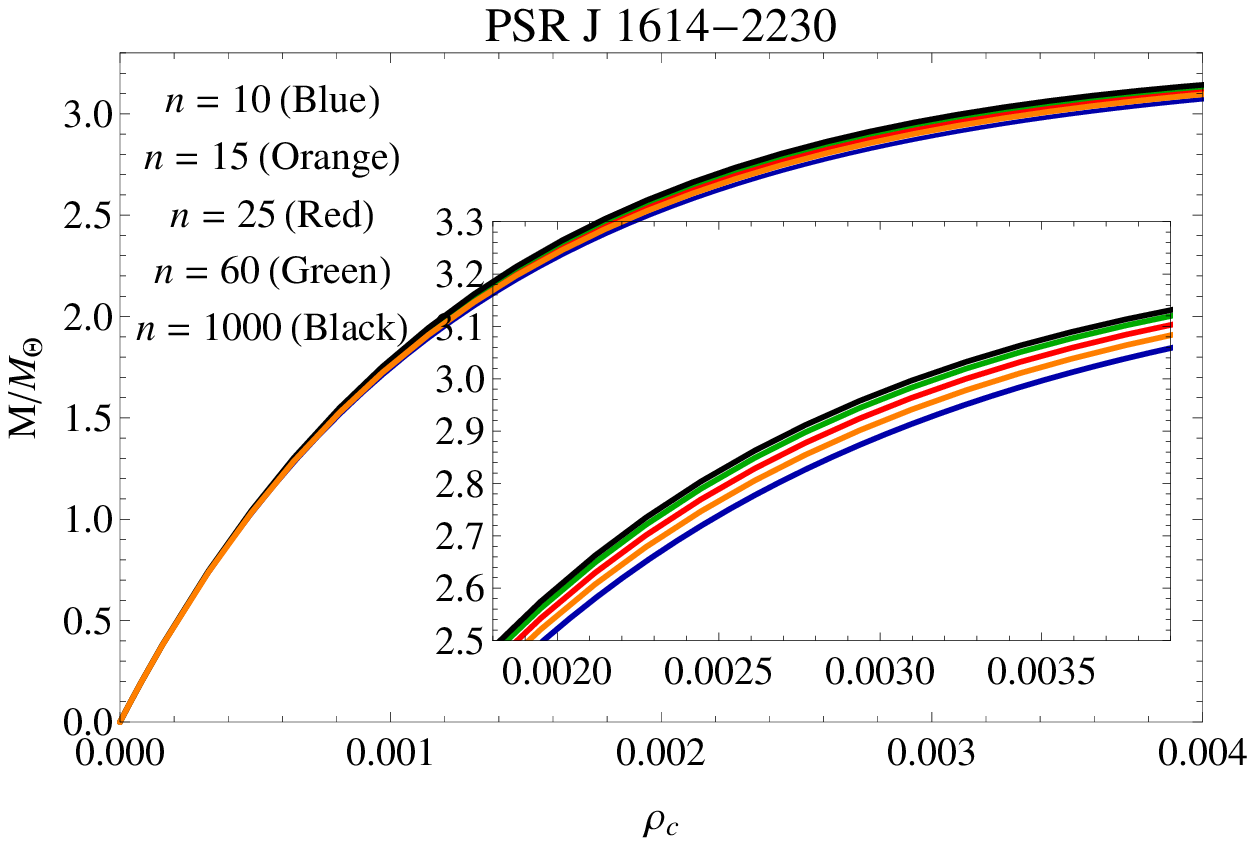}
         \includegraphics[scale=.6]{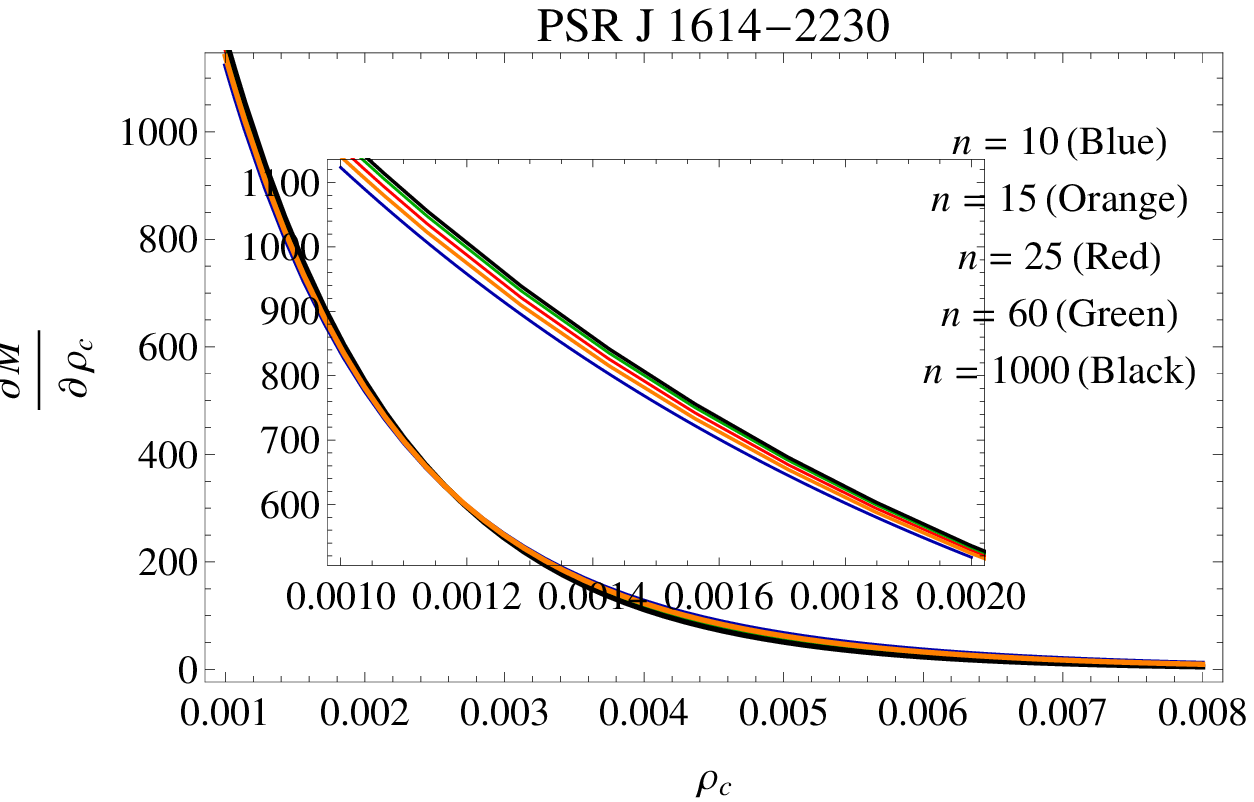}
       \caption{(Top) The variation of the mass function and (bottom) $\frac{\partial M}{\partial \rho_c}$ are plotted with respect to central density $\rho_c$ inside the stellar interior for the compact star PSR J1614-2230 for different values of `n' mentioned in the figures.}
    \label{tt}
\end{figure}

\section{Stability Analysis}\label{stab}

\subsection{Harrison-Zeldovich-Novikov stability criterion}

A stability condition for the model of compact star proposed by \cite{harri} and \cite{novi} depending on the mass and central density of the star. They suggested that for stable configuration $\frac{\partial M}{\partial \rho_c}>0$, otherwise the system will be unstable. Where $M,\,\rho_c$ denotes the mass and central density of the compact star.\\
For our present model,
\begin{eqnarray}
\frac{\partial M}{\partial \rho_c}=\frac{4}{3}\pi r_b^3 \left(1+\frac{\kappa}{3n}\rho_c r_b^2\right)^{-n-1}.
\end{eqnarray}
Above expression of $\frac{\partial M}{\partial \rho_c}$ is positive and hence the stability condition is well satisfied. The variation of the mass function and $\frac{\partial M}{\partial \rho_c}$ with respect to the central density are depicted in Fig.~\ref{tt}.

\subsection{Stability under three forces}
The stability of our present model under three different forces $viz$ gravitational force, hydrostatics force and anisotropic force can be described by the equation,
\begin{equation}\label{tov1}
-\frac{M_G(\rho+p_r)}{r^{2}}\frac{B}{A}-\frac{dp_r}{dr}+\frac{2}{r}(p_t-p_r)=0,
\end{equation}
known as TOV equation.\\
where $M_G(r) $ represents the gravitational mass within the radius $r$, which can be derived from the Tolman-Whittaker
formula and the Einstein's field equations and is defined by
\begin{equation}\label{tov2}
M_G(r)=r^{2}\frac{A'}{B}.
\end{equation}
Plugging the value of $M_G(r)$ in equation (\ref{tov1}),
this equation can be rewritten as,
\begin{equation}
F_g+F_h+F_a=0,
\end{equation}

\begin{figure}[htbp]
    \centering
        \includegraphics[scale=.55]{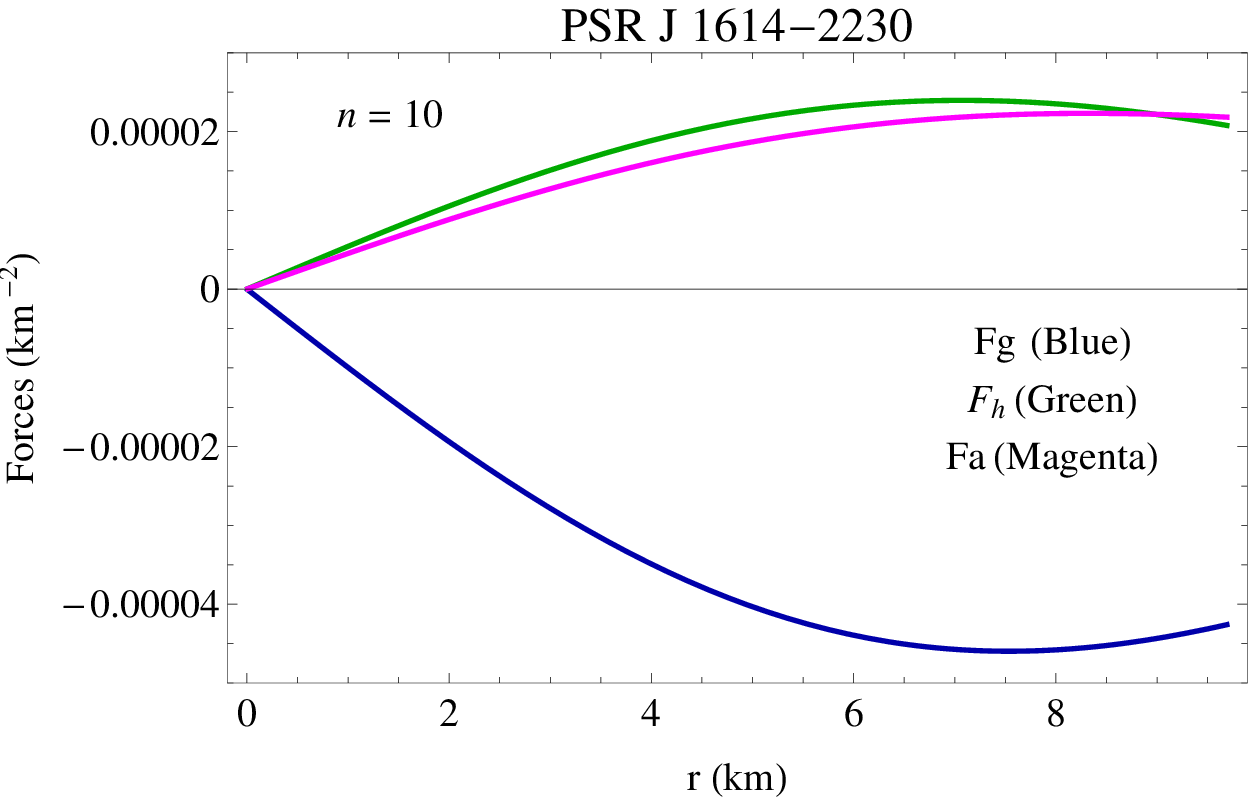}
        \includegraphics[scale=.55]{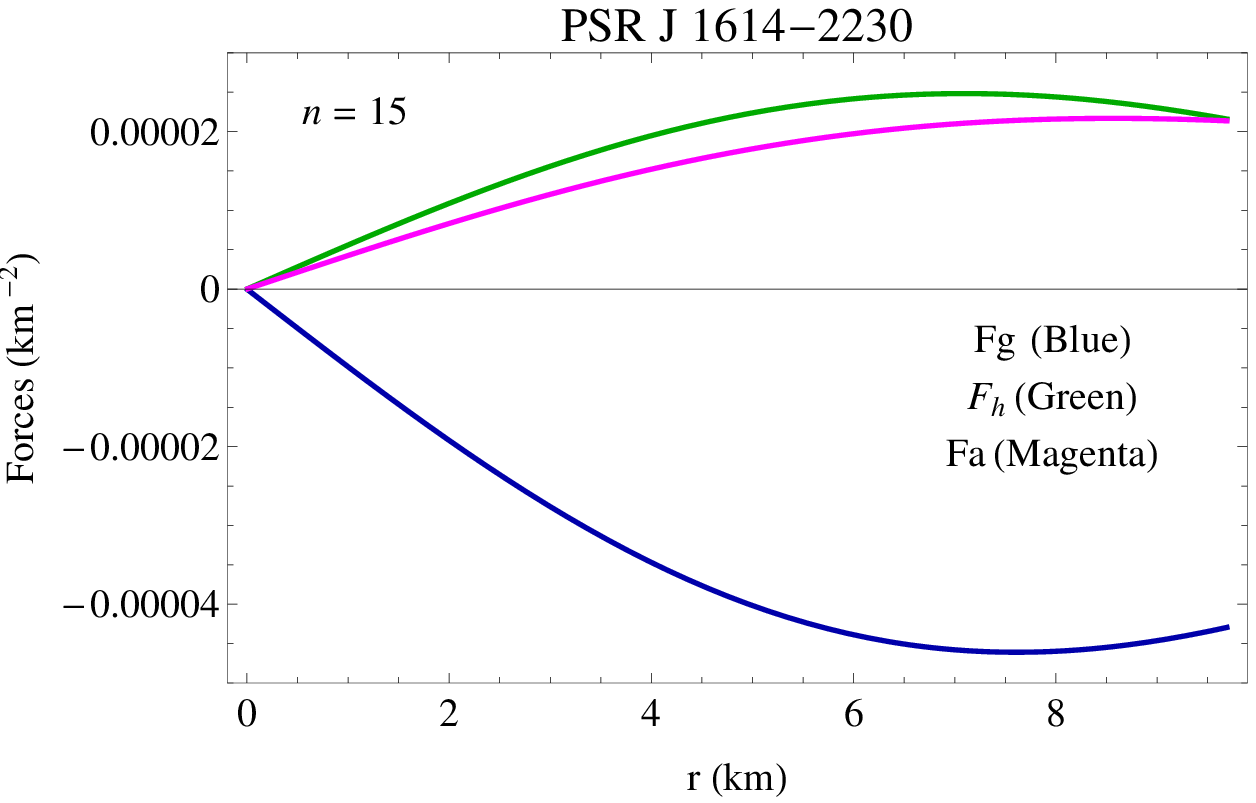}
        \includegraphics[scale=.55]{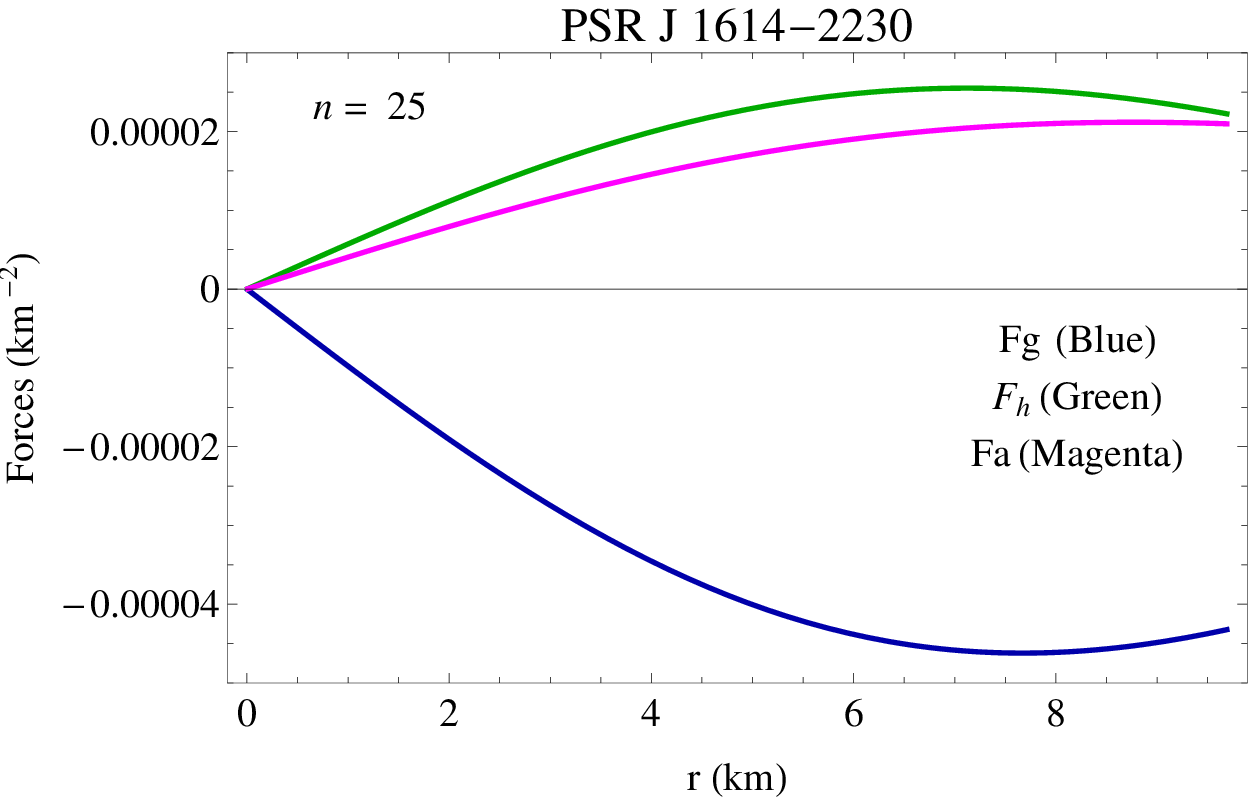}
        \includegraphics[scale=.55]{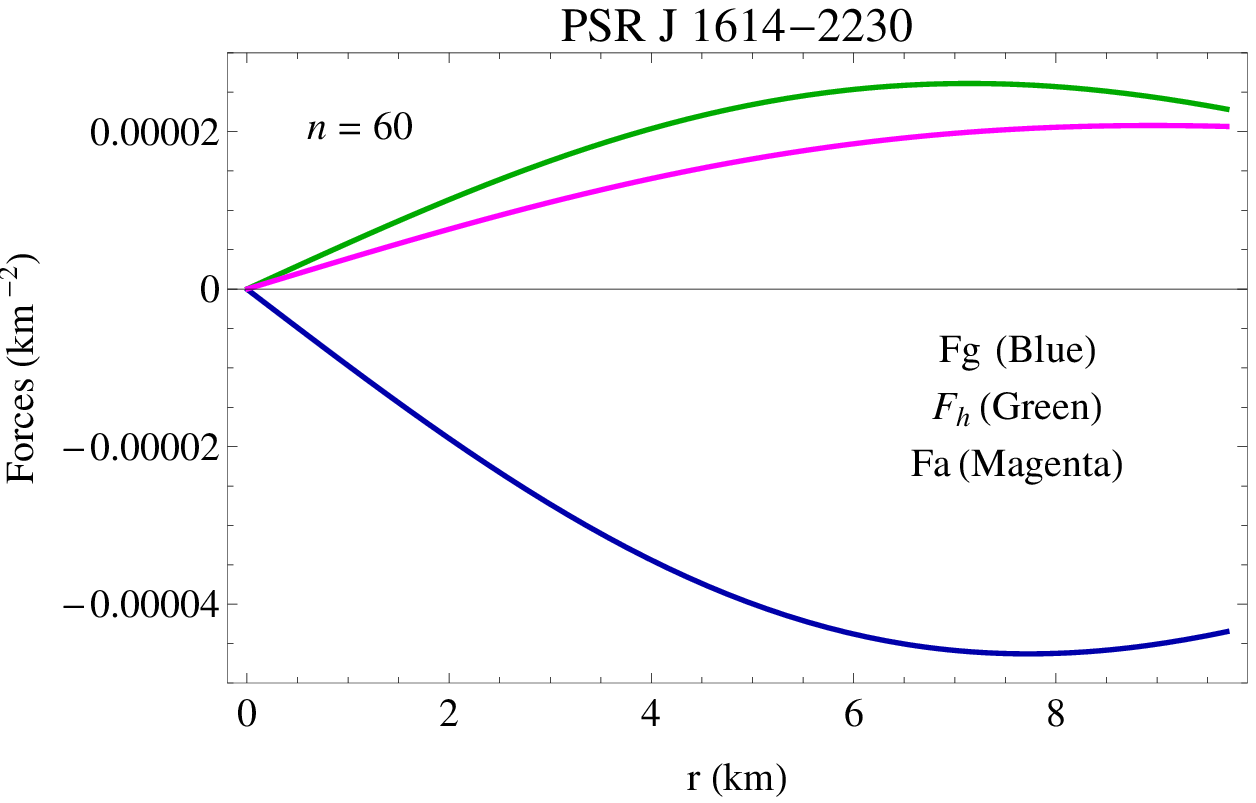}
        \includegraphics[scale=.55]{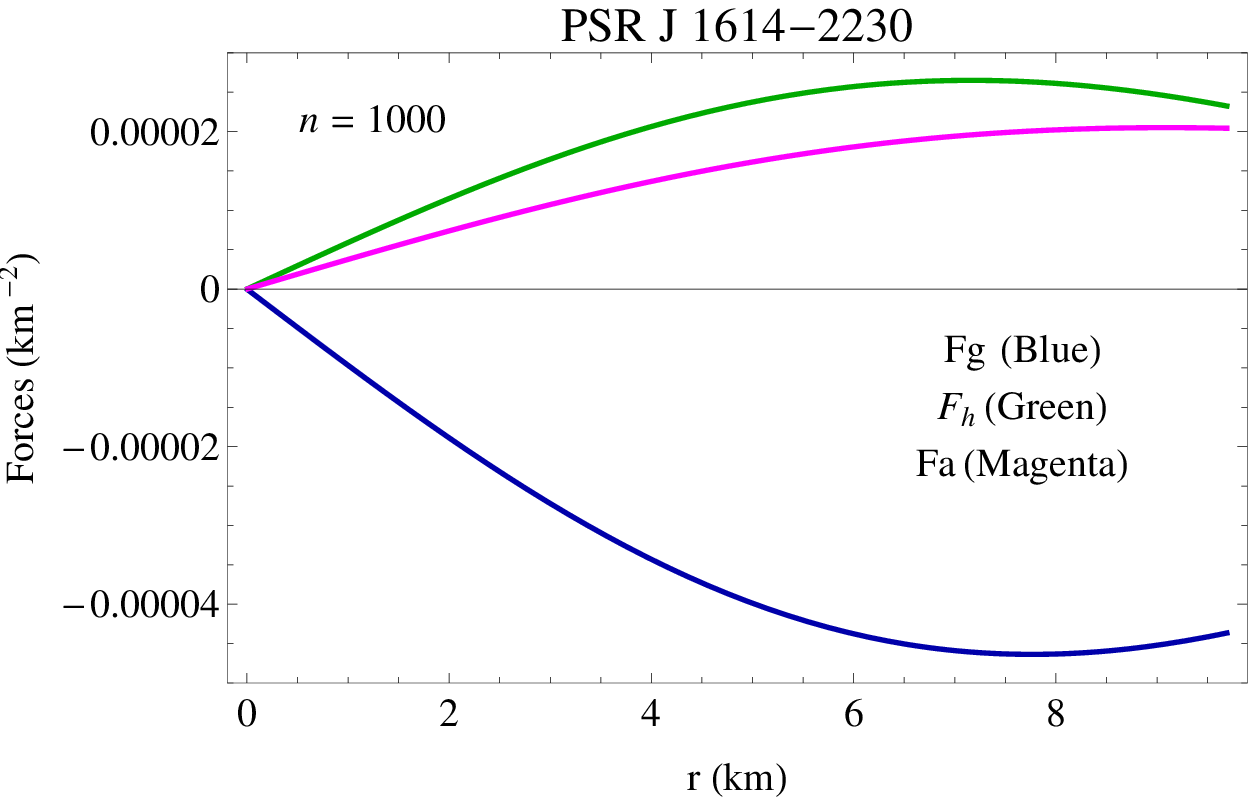}
       \caption{Gravitational, hydrostatics and anisotropic forces are plotted against $r$ inside the stellar interior for the compact star PSR J1614-2230 for different values of `n' mentioned in the figures.\label{tov}}
\end{figure}
where,
\begin{eqnarray}
F_g&=&-\frac{ C (2 + n) r \left(1 + \frac{r^2}{R^2}\right)^{-n} (r^2 + R^2)^{-1 + \frac{n}
  {2}}\Psi_1(r) }{4 \pi},\nonumber\\
  \\
  F_h&=&\frac{1}{4\pi r^3}\bigg[-1 + \left(1 + \frac{r^2}{R^2}\right)^{-n} \nonumber\\&&+ \frac{
 r^2 \left(1 + \frac{r^2}{R^2}\right)^{-n} \xi(r)}{(r^2 +
    R^2) \Big(D (2 + n) + C (r^2 + R^2)^{1 + \frac{n}{2}}\Big)^2}\bigg],\nonumber\\
    \\
    F_a&=&\frac{1 - \frac{\left(1 + \frac{r^2}{R^2}\right)^{-n} \big((1 + n) r^2 + R^2\big)}{r^2 + R^2}}{4 \pi r^2}.
\end{eqnarray}
and, \begin{eqnarray*}\xi (r)&=&D^2 n (2 + n)^2 +
    C D n (2 + n) (r^2 + R^2)^{\frac{n}{2}}\Phi\\&&+
    C^2 (r^2 + R^2)^{1 + n}\big[(4 + n (7 + 2 n)) r^2 + n R^2\big],\\
    \Phi&=&(4 + n) r^2 + 2 R^2 .\end{eqnarray*}

The three different forces acting on the system are shown in Fig.~\ref{tov} for different values of `n'. The figures show that gravitational force is negative and dominating in nature which is counterbalanced by the combine effect of hydrostatics and anisotropic forces to keep the system in equilibrium.

\subsection{Causality condition and method of cracking}
For a physically acceptable model the radial and transverse velocity of sound should lie in the range $V_{r}^{2},\,V_{t}^{2}~\in~[0,\,1]$, known as causality condition. Where radial $(V_{r}^{2})$ and transverse velocity $(V_{t}^{2})$ of sound is defined as,
\begin{eqnarray}
  V_{r}^{2} = \frac{p_r'}{\rho'}, ~~
  V_{t}^{2}= \frac{p_t'}{\rho'}.
\end{eqnarray}
The expressions for $\rho',\,p_r',\,p_t'$ have been given in eqns. (\ref{rhod})-(\ref{ptd}).
\begin{figure}[htbp]
    \centering
        \includegraphics[scale=.6]{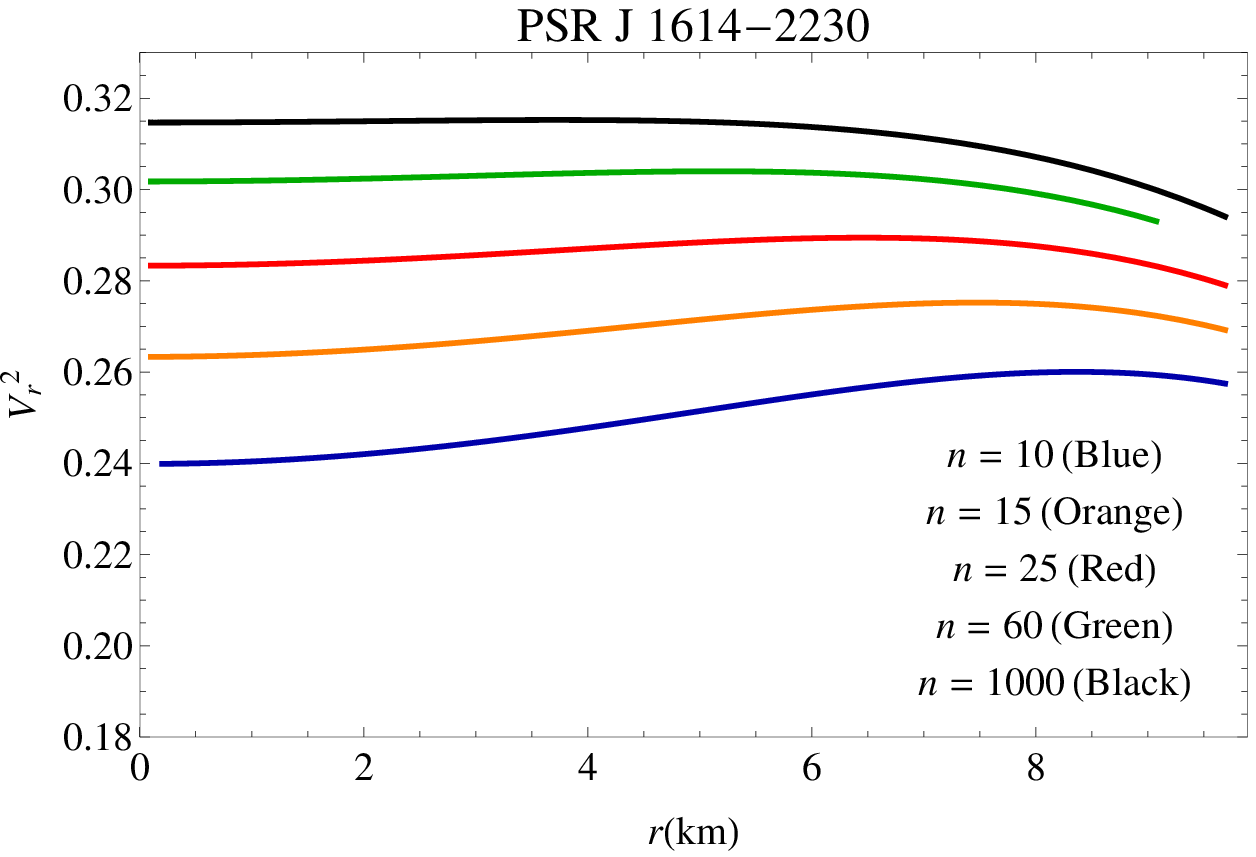}
        \includegraphics[scale=.6]{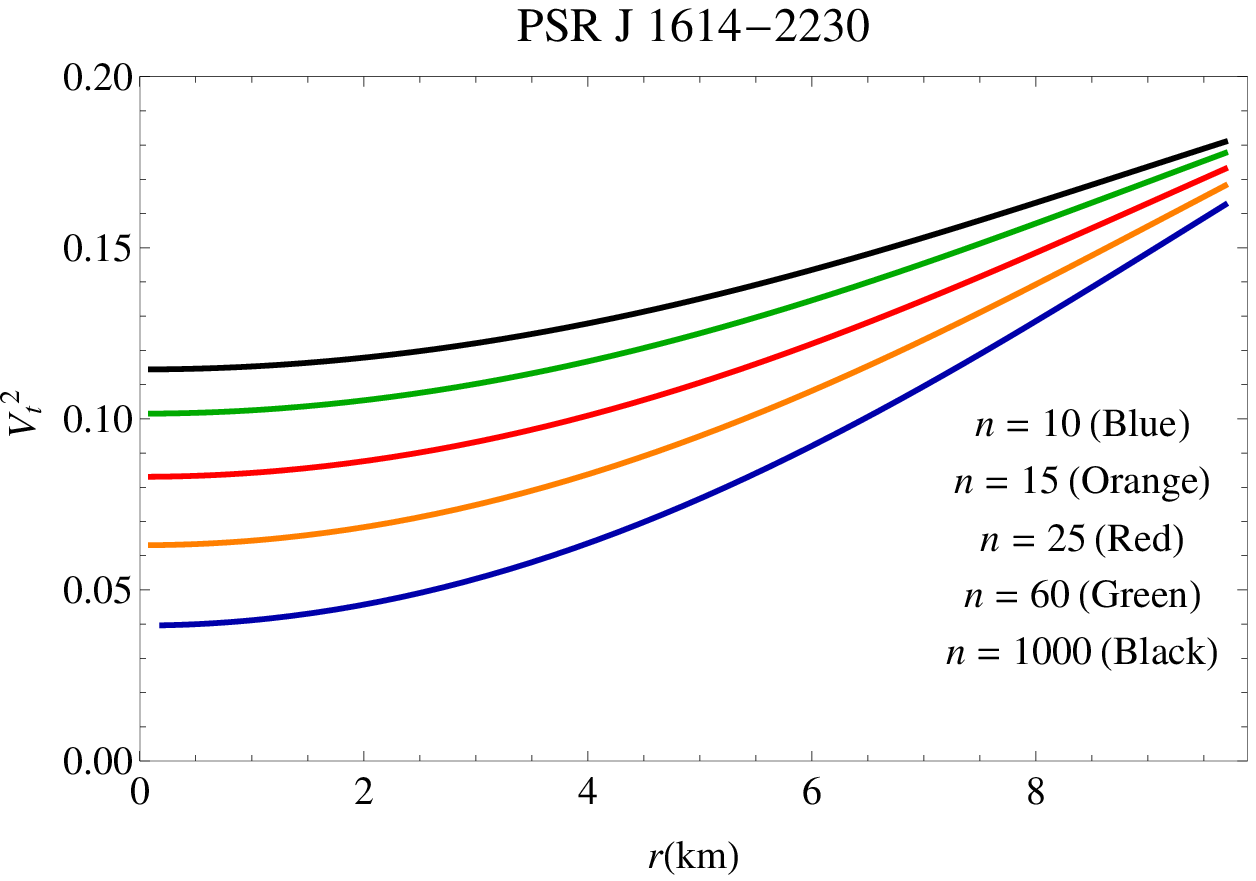}
        \includegraphics[scale=.6]{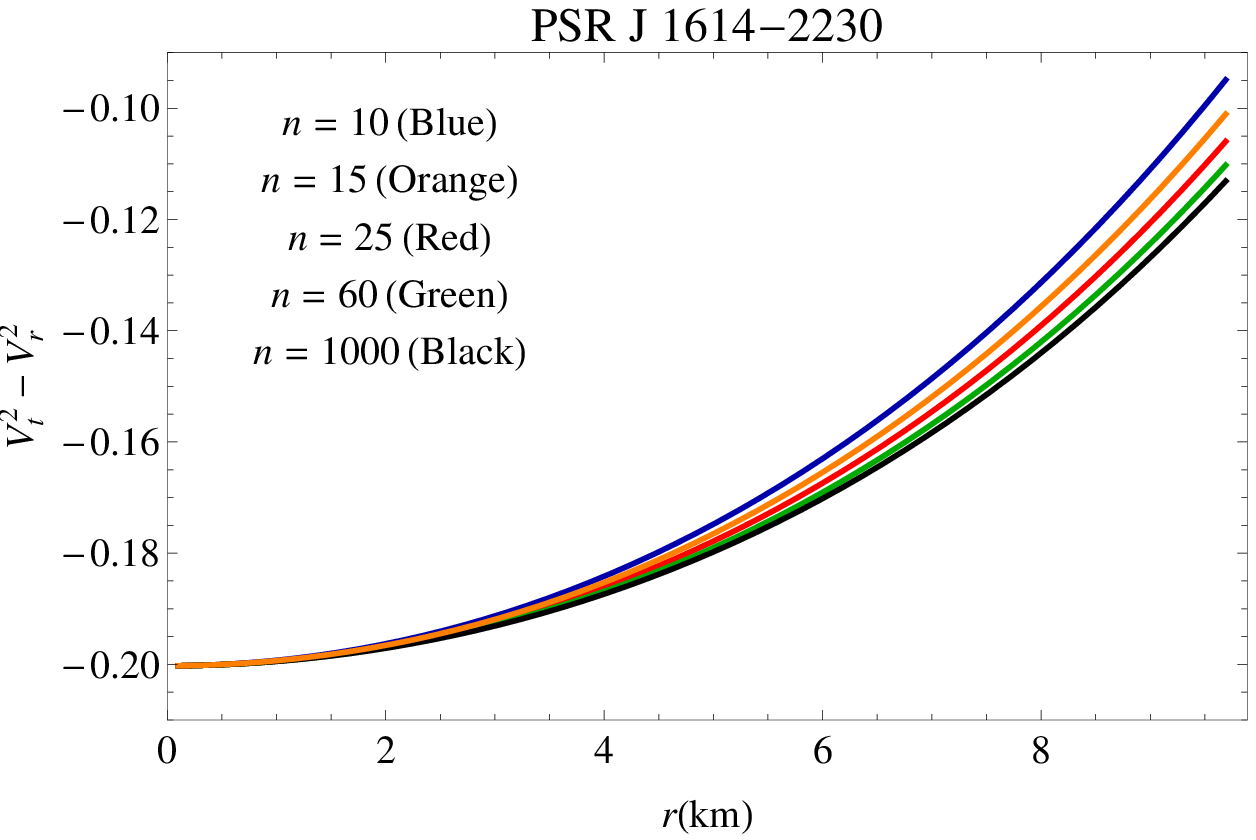}
       \caption{(Top) The square of the radial velocity $(V_r^2)$, (middle) the square of the transverse velocity $(V_t^2)$ and (bottom) the stability factor $V_t^2-V_r^2$ are plotted against $r$ inside the stellar interior for the compact star PSR J1614-2230 for different values of `n' mentioned in the figures.\label{sv}}
\end{figure}
Now, due to the complexity of the expression it is very difficult to verify the causality condition analytically. The profiles of $V_r^2$ and $V_t^2$ are shown in Fig.~\ref{sv}. It is clear from the figure that both $V_r^2$ and $V_t^2$ lie in the reasonable range. So it can be concluded that the causality condition is well satisfied. Next we want to concentrate to the stability factor of the present model which is defined as $V_t^2-V_r^2$. For the stability of a compact star model, \cite{her11} proposed the method of `cracking' and using this method, \cite{abreu11} proposed that
for a potentially stable configuration, $V_t^2-V_r^2 < 0$. From Fig.~\ref{sv} (bottom panel), we see that the stability factor is negative and hence we conclude that our model is potentially stable everywhere within the stellar interior.

\subsection{Adiabatic index}
For a particular stellar configuration, \cite{bondi645}
examined that a Newtonian isotropic sphere will be in equilibrium
if the adiabatic index $\Gamma > 4/3$ and it
gets modified for a relativistic anisotropic fluid sphere.
Based on these results,
the stability of a anisotropic stellar configuration depends on the adiabatic index $\Gamma_r$ is given by,
\begin{eqnarray}
  \Gamma_r &=& \frac{\rho+p_r}{p_r}\frac{dp_r}{d\rho},\nonumber\\
  &=&\frac{2 r^2 \Big(D n (2 + n) + 2 C (1 + n) \chi^{1 + \frac{n}{2}}\Big)}{\chi \Big(D (2 + n) \big(1- \left(1 + \frac{r^2}{R^2}\right)^n\big) +
   C \chi^{\frac{n}{2}} \Psi_2(r)\Big)}V_r^2.\nonumber\\
\end{eqnarray}
where,
\begin{eqnarray*}
\Psi_2(r)&=&(5 + 2 n) r^2+R^2-\left(1 + \frac{r^2}{R^2}\right)^n (r^2 + R^2),\\
\chi&=& r^2 + R^2,
\end{eqnarray*}
and the expression of $\frac{dp_r}{d\rho}$ has been given in previous subsection.
\begin{figure}[htbp]
    \centering
        \includegraphics[scale=.6]{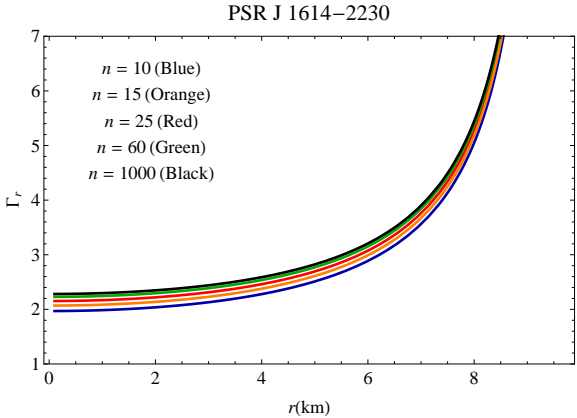}
       \caption{$\Gamma_r$ is plotted against $r$ inside the stellar interior for the compact star PSR J1614-2230 for different values of `n' mentioned in the figure.\label{gama}}
\end{figure}

For our present model, the profile of $\Gamma_r$ are shown in Fig.~\ref{gama} for different values of `n'. From the figure, we see that the profile of radial adiabatic index are monotonic increasing function of $r$ and $\Gamma_r >4/3$ every where within the stellar configuration and hence the condition of stability is satisfied.

\section{Discussion}\label{dis}
Present paper provides a new generalized model of compact star by assuming a physically reasonable metric potential together with a pressure anisotropy. We matched our interior solution to the exterior Schwarzschild line element at the boundary to fix the values of the different constants. From the boundary conditions we have obtained the values of $R,\,C,\,D$ for the compact star PSR J1614-2230 and EXO 1785-248 with masses $1.97~M_{\odot}$, $1.3~M_{\odot}$ respectively and radii $9.69~$km, $8.85~$ km respectively in table~\ref{tab1} and table~\ref{tab2} for different values of the dimensionless parameter `n'. From the tables, it is clear that the numerical values of the constants $C$ and $D$ decreases with increasing `n' where as the numerical values of $R$ increases with `n' increasing. One notable thing is that the numerical values of $C$ decreases rapidly whereas the numerical value of $D$ decreases steadily. All the profiles are depicted for the compact star  PSR J1614-2230. We have plotted the profiles of matter density ($\rho$), radial pressure ($p_r$) and transverse pressure ($p_t$) for different values of `n' and we see that all are positive and monotonic decreasing function of $r$. The values of the model parameters like central density, surface density, central pressure, surface transverse pressure, central value of radial adiabatic index for the above mentioned two stars are presented in table~\ref{tab3} and table~\ref{tab4}. We also note that the cental values of pressure and density decreases with increasing `n', it is evident from table~II as well as from Fig.~\ref{rho}, Fig.~\ref{pr}. On the other hand, the surface density of the star increases as `n' increases. Plugging $G$ and $c$ in the expression of $\rho$ and $p_r$, we obtained the central density and central pressure for different values of `n' which lies in the range $1.57\times 10^{15}~gm/cm^3-1.64\times 10^{15}~gm/cm^3$ and $2.05\times 10^{35}~dyne/cm^2-2.26\times 10^{35}~dyne/cm^2$ respectively. It can also be noted that, the surface density lies in the range $7.43\times 10^{14}~gm/cm^3-7.62\times 10^{14}~gm/cm^3$ for the compact star PSR J1614-2230. Also, the central density and central pressure for different values of `n' lies in the range $1.36\times 10^{15}~gm/cm^3-1.32\times 10^{15}~gm/cm^3$ and $8.802\times 10^{34}~dyne/cm^2-1.005\times 10^{35}~dyne/cm^2$ respectively and surface density lies in the range $6.96\times 10^{14}~gm/cm^3-7.14\times 10^{14}~gm/cm^3$ for the compact star EXO 1785-248. From the figure, it is clear that the transverse pressure $p_t$ always dominates the radial pressure $p_r$ and it creates a positive pressure anisotropy and hence repulsive force towards the boundary. With the help of graphical representation, we have shown that our model satisfies all the energy conditions and $(p_r+2p_t)/\rho$ is monotonic decreasing and less than $1$. The radial adiabatic index $\Gamma_r > 4/3$ and the causality conditions are satisfied by our model. The stability conditions of the model have been tasted by different conditions. The equation of state parameter $\omega_r$ is monotonic decreasing function of `r' but $\omega_t$ is monotonic increasing. Both of them and lies in the range $0 < \omega_r,\,\omega_t < 1$ (Fig.~\ref{omega}). The forces acting on the present model is depicted in Fig.~\ref{tov} and it shows the effect of gravitational force ($F_g$) and anisotropic force ($F_a$) are increased with the increasing value of `n'. The central values of radial adiabatic index for the compact star PSR J1614-2230 and EXO 1785-248  are obtained in table.~\ref{tab3} and table.~\ref{tab4}. We see that the cental values of radial adiabatic index increases as `n' increases. So one can conclude that the increasing value of `n' makes the system more stable in respect of the test of adiabatic index. One can also note that the central values of radial and transverse velocity of sound increase with the increasing values of `n'. To check the behavior of the radial and transverse pressure with the matter density we draw the profiles of $p_r$ versus $\rho$ and $p_t$ versus $\rho$ in Fig.~\ref{drho}. The potential stability condition of the present model is also satisfied.
So we can conclude that solution obtained in this
paper can be used as a successful model for the description of ultra-compact stars.


\section*{ Acknowledgements}

P.B is thankful to IUCAA, Government of India, for providing visiting associateship.

\end{document}